# Liquid State Machine Learning for Resource and Cache Management in LTE-U Unmanned Aerial Vehicle (UAV) Networks

Mingzhe Chen*, Walid Saad†, and Changchuan Yin*

*Beijing Laboratory of Advanced Information Network, Beijing University of Posts and Telecommunications, Beijing, China 100876, Emails: chenmingzhe@bupt.edu.cn, ccyin@ieee.org.

†Wireless@VT, Bradley Department of Electrical and Computer Engineering, Virginia Tech, Blacksburg, VA, USA, Email: walids@vt.edu.

**Abstract**

In this paper, the problem of joint caching and resource allocation is investigated for a network of cache-enabled unmanned aerial vehicles (UAVs) that service wireless ground users over the LTE licensed and unlicensed (LTE-U) bands. The considered model focuses on users that can access both licensed and unlicensed bands while receiving contents from either the cache units at the UAVs directly or via content server-UAV-user links. This problem is formulated as an optimization problem which jointly incorporates user association, spectrum allocation, and content caching. To solve this problem, a distributed algorithm based on the machine learning framework of *liquid state machine* (LSM) is proposed. Using the proposed LSM algorithm, the cloud can predict the users' content request distribution while having only limited information on the network's and users' states. The proposed algorithm also enables the UAVs to autonomously choose the optimal resource allocation strategies that maximize the number of users with stable queues depending on the network states. Based on the users' association and content request distributions, the optimal contents that need to be cached at UAVs as well as the optimal resource allocation are derived. Simulation results using real datasets show that the proposed approach yields up to $33.3\%$ and $50.3\%$ gains, respectively, in terms of the number of users that have stable queues compared to two baseline algorithms: Q-learning with cache and Q-learning without cache. The results also show that LSM significantly improves the convergence time of up to $33.3\%$ compared to conventional learning algorithms such as Q-learning.

# I. Introduction

The use of aerial wireless communication platforms carried by unmanned aerial vehicles (UAVs) is seen as a promising approach to improve the coverage and capacity of future wireless networks [2]. UAVs can have three key functions: aerial base stations, aerial relays, and user equipments. Due to their flying nature, UAVs can be used as aerial base stations to provide line-of-sight (LoS) connections toward ground users and, thus, potentially improving the rate, delay, and overall performance of wireless networks. Moreover, UAVs can be used as user equipments [3] connected to the cellular network for information dissemination and data collection. In addition, UAVs can act as relays between a source and a destination for scenarios in which a LoS link does not exist or for backhaul connectivity of base stations. However, deploying UAVs for wireless communication purposes faces many challenges [4]–[15] that include air-to-ground channel modeling, optimal deployment, energy efficiency, path planning, resource management, and security.

## *A. Related Work*

The existing literature in [4]–[15] has studied a number of problems related to UAV communications and networking. For instance, in [4]–[6], the deployment of an unmanned aerial vehicle acting as a wireless base station that provide coverage for ground users is analyzed. The authors in [7] proposed a statistical propagation model for predicting the air-to-ground path loss between a low altitude platform and a ground user. In [8], the authors investigated a multi-tier drone-enabled network complementing a terrestrial heterogeneous network. The work in [9] studied the use of LTE-Unlicensed (LTE-U) technology to enhance the achievable broadband throughput of UAVs. The authors in [10] studied the behavior of air-to-ground mmWave bands for UAVs using ray tracing simulations. In [11], the authors investigated the problem of efficient power control in UAV-supported ultra dense networks. The work in [12] studied the throughput maximization problem in mobile relaying systems by optimizing the source/relay transmit power along with the relay trajectory. Meanwhile, in [13], the authors jointly considered the radio resource allocation, three-dimensional placement, and user association of UAVs in the internet of things networks. The authors in [14] provided a comprehensive survey on available air-to-ground channels for UAV communication. In [15], the authors developed a new framework to optimize the minimum throughput of ground users by optimizing the multiuser communication scheduling and association jointly with the UAVs' trajectory and power control. However, most of these existing works [4]–[15] are focused on the direct communication links between UAVs and ground users and do not account for the ground-to-air fronthaul communications between the core network and the UAVs. Indeed, capacity-constrained fronthaul links will significantly limit the transmission rate of the links from the UAVs to the ground users. In order to reduce the traffic load on the fronthaul links and improve the performance of UAV-based communication, one promising approach is to cache popular content [16]–[18] at the UAVs thus allowing them to directly transmit data to the ground users without using wireless fronthaul transmissions, whenever possible [19]–[21].

To further improve performance and overcome the spectrum scarcity problem, the UAVs can be also equipped with LTE over the unlicensed band capabilities thus allowing them to use both licensed and unlicensed spectrum to service their ground users. Recently, there has been significant interest in studying the performance of LTE-U enabled cellular networks such as in [9], [22]–[27]. However, the LTE-U works in [22]–[27] do not consider the use of LTE-U with UAV-carried base stations. Meanwhile, even though the work in [9] considers LTE-U resource allocation in a UAV network, this work does not consider the effect of the limited-capacity cloud-UAVs links. In addition, the work in [9] does not exploit the use of caching at the UAV side. Caching the popular contents at UAVs can help overcome the limited capacity of the cloud-UAVs links.

### *B. Contributions*

The main contribution of this paper is a novel resource allocation framework that allows cache-enabled UAVs to effectively service ground users over licensed and unlicensed bands in a cloud network under fronthaul capacity constraints. To our best knowledge, *this is the first work to jointly consider the joint use of the caching and LTE-U for UAV-assisted communication*. Our main contributions include:

- We *jointly* analyze the problems of resource allocation over licensed and unlicensed bands, user association, and caching content replacement. We formulate the problem as an optimization problem whose goal is to maximize the number of the stable queue users.
- To solve this resource allocation problem, we propose a novel self-organizing, decentralized algorithm to maximize the number of stable queue users. Unlike previous studies [4], [6]– [15], that overlook the limited-capacity of the UAV-cloud links, we propose a novel learning approach based on the powerful framework of *liquid state machine (LSM)* [28] that can jointly perform caching and resource allocation in a network of cache-enabled LTE-U UAVs. The use of LSM enables the cloud to quickly learn the users' content request distribution so as to determine the content caching strategy for each UAV. It also enables the UAVs to autonomously adjust their spectrum allocation schemes to service users over licensed and unlicensed bands.

- Within the proposed LSM algorithm, we propose a search-based algorithm that can find the optimal resource allocation and derive optimal contents to cache so as to calculate the number of stable queue users. We also prove that the proposed search-based algorithm can find the optimal resource allocation for a given user association.
- Simulation results show that the proposed approach yields gains of up to $33.3\%$ and $50.3\%$ in terms of the number of users having a stable queue compared to Q-learning with cache and Q-learning without cache, respectively.

The rest of this paper is organized as follows. The system model and problem formulation are described in Section II. The LSM-based algorithms for content request distribution prediction and resource allocation are proposed in Section III. In Section IV, numerical simulation results are presented and analyzed. Finally, conclusions are drawn in Section V.

## II. System Model and Problem Formulation

Consider the downlink of an LTE-U network composed of a set $\mathcal{K}$ of $K$ UAVs that service a set $\mathcal{U}$ of $U$ LTE-U users and $W$ WiFi access points (WAPs) that only service $N_w$ WiFi users, as shown in Fig. 1. In this model, the UAVs are equipped with cache storage units and can be deployed to act as flying cache-enabled LTE-U base stations to serve a set $\mathcal{U}$ of $U$ ground users. The UAVs are controlled by a cloud-based server. Here, we consider *dual mode* cache-enabled UAVs that are able to access both the licensed and unlicensed bands. Each UAV can allocate at most one type of resource (i.e., licensed or unlicensed resources) to each user. The transmissions from the cloud to the UAVs occur over wireless fronthaul links using the licensed cellular band.

In this system, a frequency division duplexing (FDD) mode is considered for LTE on the licensed band. The FDD mode separates the licensed band for the downlink LTE-U users. A time division duplexing (TDD) mode with duty cycle method [29] is considered for LTE-U.

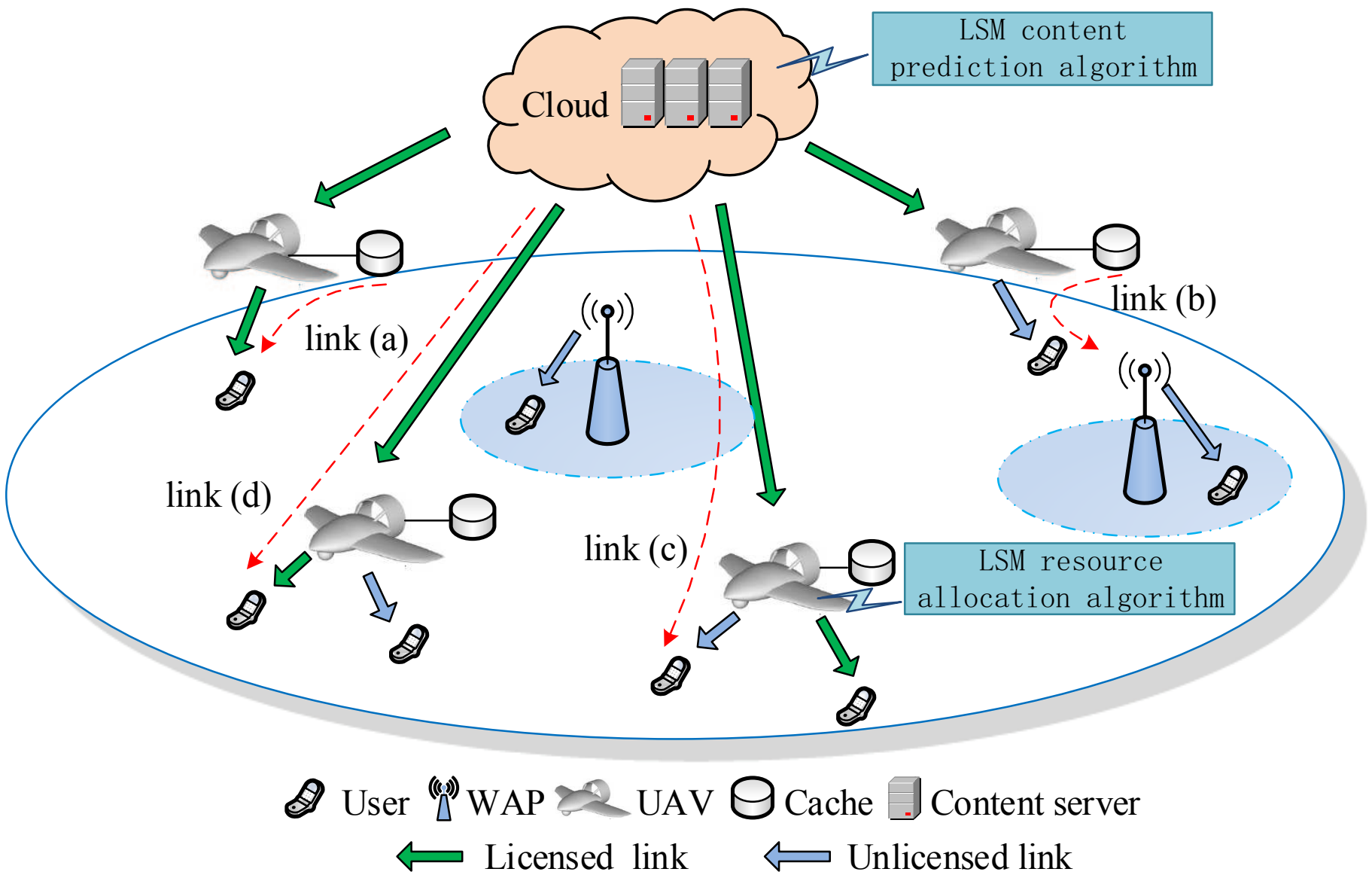

Fig. 1. The architecture of a network of cache-enabled LTE-U UAVs.

Using the duty cycle method, the UAVs can use a discontinuous, duty-cycle transmission pattern so as to guarantee the transmission rate of WiFi users. In this case, the unlicensed band time slots will be divided between LTE-U and WiFi users. In particular, LTE-U transmits for a fraction $\vartheta$ of time and will be muted for a fraction $1-\vartheta$ of time which is allocated for WiFi transmission. The WAPs transmit using a standard carrier sense multiple access with collision avoidance (CSMA/CA) protocol and its corresponding RTS/CTS access mechanism [30].

In our model, all ground users will request contents of equal size $L$ from a set $\mathcal{N}$ of $N$ contents that are stored at a cloud-based content server. The content request distribution of each user is $\boldsymbol{p}_i = [p_{i1}, \ldots, p_{iN}]$, where $p_{ij}$ represents the probability that user $i$ requests content $j$. This distribution can be obtained by the machine learning algorithm that we will detail in Section III. Each UAV $k$ is equipped with a storage unit that can store a set $\mathcal{C}_k$ of $C$ popular contents that the users can request. If properly done, caching at the UAVs can significantly offload the fronthaul traffic of UAVs since each UAV can directly transmit its stored contents to the users without using fronthaul links. Hereinafter, caching at the UAVs is referred to as "UAV cache". The cached contents at a UAV are assumed to be periodically refreshed at off peak hours when the UAVs return to their docking stations. Table I provides a summary of the notations used hereinafter.

TABLE I
LIST OF NOTATIONS

| Notation | Description | Notation | Description |
|---|---|---|---|
| $U$ | Number of LTE-U users | $\vartheta$ | Fraction of WiFi time slots allocated to UAVs |
| $K$ | Number of UAVs | $F_l, F_u$ | Bandwidth of licensed band and unlicensed band |
| $W$ | Number of WAPs | $T_{\rm W}$ | Number of WiFi time slots constructed an LTE time slot |
| $N_w$ | Number of WiFi users | $l_{ki}^{\rm LoS}$, $l_{ki}^{\rm NLoS}$ | Path loss of LoS/NLoS link from UAV $k$ to user $i$ |
| $F_C$ | Bandwidth of fronthaul | $\mathcal{U}_k$ | The set of the users associated with UAV $k$ |
| $L$ | Size of each content | $\boldsymbol{u}, \boldsymbol{e}$ | Resource allocation vector |
| $N$ | Number of contents at cloud | $\boldsymbol{x}_i$, $\boldsymbol{m}_k$ | Input of LSM algorithm |
| $\boldsymbol{p}_i$ | Content request distribution of user $i$ | $\boldsymbol{y}_i$, $\boldsymbol{b}_k$ | Output of LSM algorithm |
| $C$ | Number of contents stored at each UAV | $N_W$ | Number of neurons in LSM |
| $\mathcal{C}_k$ | The set of contents stored at UAV $k$ | $\bar{l}_{ki}^{l}$ | Average licensed path loss from UAV $k$ to user $i$ |
| $R_{Ck}$ | Rate from the cloud to UAV $k$ | $\mathbb{P}_{ij}$ | Probability connection between neurons $i$ and $j$ |
| $R_{\rm W}$ | Data rate of a WiFi user | $\bar{l}_{ki}^{u}$ | Average unlicensed path loss from UAV $k$ to user $i$ |
| $Q_i\left(t\right)$ | Queue of user $i$ at time $t$ | $U_C\left(t\right)$ | Number of users that request contents from cloud |
| $v_j\left(t\right)$ | Action state of neuron $j$ | $R_{lki}$ | Rate of user $i$ associated with UAV $k$ via licensed band |
| $\boldsymbol{f}_j, \boldsymbol{f}_j^{\alpha}$ | Output function of LSM | $R_{uki}$ | Rate of user $i$ associated with UAV $k$ via unlicensed band |

## A. WiFi data rate analysis

For the WiFi network, we assume that the WAPs will adopt a CSMA/CA scheme with binary slotted exponential backoff. Therefore, the saturation capacity of $N_w$ users sharing the same unlicensed band can be expressed by [10]:

$$R\left(N_w\right) = \frac{P_{\rm tr}\left(N_w\right) P_{\rm s}\left(N_w\right) E\left[A\right]}{(1-P_{\rm tr}\left(N_w\right))T_\sigma + P_{\rm tr}(N_w)P_{\rm s}(N_w)T_{\rm s} + P_{\rm tr}(N_w)(1-P_{\rm s}(N_w))T_{\rm c}}, \tag{1}$$

where $P_{\rm tr}\left(N_w\right) = 1-\left(1-\tau\right)^{N_w}$ with $P_{\rm tr}\left(N_w\right)$ being the probability that there is at least one transmission in a time slot and $\tau$ being the transmission probability of each user. $P_{\rm s}\left(N_w\right) = N_w\tau\left(1-\tau\right)^{N_w-1}/P_{\rm tr}\left(N_w\right)$, is the successful transmission probability, $T_s$ is the average time that the channel is sensed busy because of a successful transmission, $T_c$ is the average time that the channel is sensed busy by each station during a collision, $T_\sigma$ is the duration of an unoccupied slot time, and $E\left[A\right]$ is the average packet size. In our model, the WiFi network adopts conventional distributed coordination function access and RTS/CTS access mechanisms. $T_{\rm c}$ and $T_{\rm s}$ are computed as done in [10].

We assume that one LTE time slot consists of $T_W$ WiFi time slots. Based on the duty cycle mechanism, the UAVs can occupy $\vartheta$ fraction of $T_W$ time slot on the unlicensed band while the WiFi users can occupy a fraction $(1-\vartheta)$ fraction of $T_W$ time slots as shown in Fig. 2. Thus, the per WiFi user rate can be given by:

$$R_w = \frac{R(N_w)(1-\vartheta)}{N_w}, \tag{2}$$

where $N_w$ is the number of WiFi users on the unlicensed band. Given the rate requirement of each WiFi user $\gamma$, the fraction of the time slot on the unlicensed band allocated to the LTE-U users can be given by $\vartheta \leq 1 - N_w\gamma/R(N_w)$.

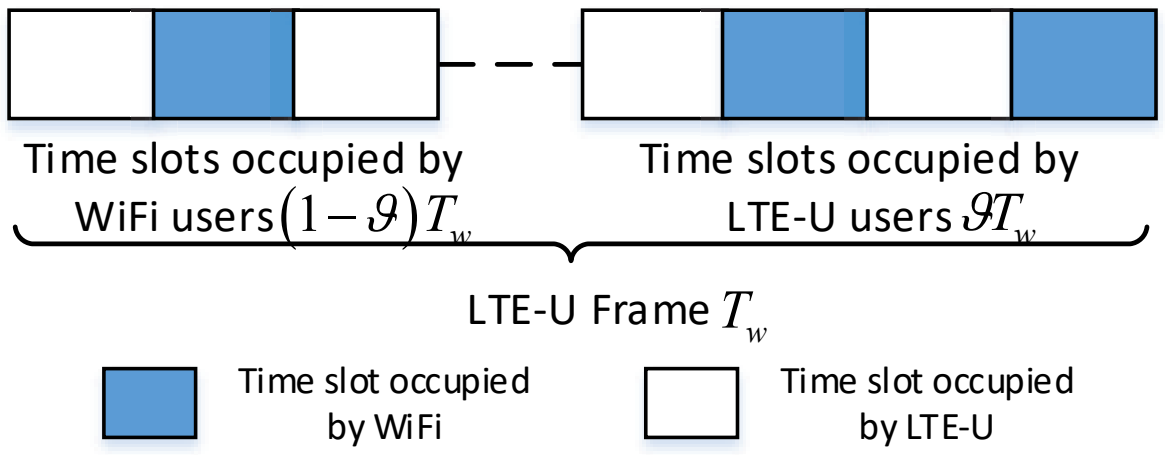

Fig. 2. Illustrative example on the time slot allocation between WiFi and LTE-U users.

### B. *UAV data rate analysis*

Next, we define the data rate of each user associated with any given UAV. The UAVs' content transmission link consists of the wireless fronthaul links that connect each UAV to the cloud (ground-to-air links) and the UAV-users links (air-to-ground links). As done in [4], [6], [7], we consider probabilistic LoS and non-line-of-sight (NLoS) links over the licensed band for both the UAVs' fronthaul links and UAV-user links. In such a model, NLoS links experience higher attenuation than LoS links due to the shadowing and diffraction loss.

*1) UAVs-users links over the licensed band:* The LoS and NLoS path loss from UAV $k$ to user $i$ will be given by (in dB) [7]:

$$l_{ki}^{\text{LoS}} = 20\log\left(\frac{4\pi d_{ki} f}{c}\right) + \eta_{\text{LoS}}^{l},$$

$$l_{ki}^{\text{NLoS}} = 20\log\left(\frac{4\pi d_{ki} f}{c}\right) + \eta_{\text{NLoS}}^{l},$$

where $20\log(d_{ki} f 4\pi/c)$ is the free space path loss with $d_{ki}$ being the distance between user $i$ and UAV $k$, $f$ being the carrier frequency, and $c$ being the speed of light. $\eta_{\text{LoS}}^{l}$ and $\eta_{\text{NLoS}}^{l}$ represent, respectively, additional attenuation factors due to the LoS/NLoS connections over the licensed band. As shown in [4], [6], [7], the probability of LoS connection depends on the environment, density and height of buildings, the locations of the user and the UAV, and the elevation angle between the user and the UAV. The LoS probability is given by [7]:

$$\Pr\left(l_{ki}^{\text{LoS}}\right) = \left(1 + X\exp\left(-Y\left[\phi_{ki} - X\right]\right)\right)^{-1}, \tag{3}$$

where $X$ and $Y$ are constants which depend on the environment (rural, urban, dense urban, or others) and $\phi_{ki} = \sin^{-1}\left(h_k/d_{ki}\right)$ is the elevation angle. Consequently, the average path loss from UAV $k$ to user $i$ will be given by [7]:

$$\bar{l}_{ki}^{l} = \Pr\left(l_{ki}^{\text{LoS}}\right) \times l_{ki}^{\text{LoS}} + \Pr\left(l_{ki}^{\text{NLoS}}\right) \times l_{ki}^{\text{NLoS}}, \tag{4}$$

where $\Pr\left(l_{ki}^{\text{NLoS}}\right) = 1 - \Pr\left(l_{ki}^{\text{LoS}}\right)$. Based on the path loss, the downlink rate of user $i$ associated with UAV $k$ on the licensed band at time $t$ will be:

$$R_{lki}(u_{ki}\left(t\right)) = u_{ki}\left(t\right) F_l \log_2\left(1 + \frac{P_K 10^{\bar{l}_{ki}^{l}/10}}{\sum\limits_{j\in\mathcal{K}, j\neq k} P_K 10^{\bar{l}_{ji}^{l}/10} + P_C h_i + \sigma^2}\right), \tag{5}$$

where $F_l$ is the downlink bandwidth on the licensed band, $P_K$ is the transmit power of each UAV, $h_i$ is the channel gain between user $i$ and the cloud, and $P_C$ is the transmit power of the cloud over the fronthaul. $\sigma^2$ is the power of the Gaussian noise. Finally, $u_{ki}\left(t\right)$ is the fraction of the downlink licensed band allocated from UAV $k$ to user $i$ at time $t$ with $\sum_i u_{ki}\left(t\right) = 1$.

*2) UAVs-users links over the unlicensed band:* In the considered model, the UAVs can only use the unlicensed band of the WiFi networks whenever the WiFi users' rate requirement is satisfied. Based on (2), we obtain a fraction $\vartheta$ of a time slot over the unlicensed band that can be occupied by UAVs. Therefore, the downlink rate of user $i$ associated with UAV $k$ on the unlicensed band can be given by:

$$R_{uki}(e_{ki}\left(t\right)) = e_{ki}(t)\,\vartheta F_u \log_2\left(1 + \frac{P_K 10^{\bar{l}_{ki}^{u}/10}}{\sum\limits_{j\in\mathcal{K}, j\neq k} P_K 10^{\bar{l}_{ji}^{u}/10} + \sigma^2}\right), \tag{6}$$

where $\bar{l}_{ki}^{u}$ is the average path loss over the unlicensed band, $F_u$ is the bandwidth of the unlicensed band, and $e_{ki}\left(t\right)$ is the fraction of $\vartheta$ with $\sum_i e_{ki}\left(t\right) = 1$.

*3) Cloud-UAVs ground-to-air links:* The LoS and NLoS path loss from the cloud to UAV $k$ can be given by [6]:

$$L_k^{\text{LoS}} = d_{Ck}^{-\beta}, \tag{7}$$

$$L_k^{\text{NLoS}} = \varsigma d_{Ck}^{-\beta}, \tag{8}$$

where $\varsigma$ is the additional path loss of the NLoS connection and $d_{Ck}$ is the distance between UAV $k$ and the cloud. The average path loss $\bar{L}_k$ of the fronthaul link of UAV $k$ can be computed using (3) and (4). Here, we assume that the total bandwidth of the UAVs' fronthaul is $F_C$ which is equally divided among the users that received the contents from the cloud. Therefore, the fronthaul rate of each user that is associated with UAV $k$ and requests the contents from the cloud is:

$$R_{Ck}\left(t\right) = \frac{F_C}{U_C\left(t\right)} \log_2\left(1 + \frac{P_C \bar{L}_k}{\sum\limits_{j \in \mathcal{K}, j \neq k} P_K 10^{\bar{l}_{ki}/10} + \sigma^2}\right), \tag{9}$$

where $U_C(t)$ is the number of the users that receive a content from the cloud at time $t$. $U_C(t)$ can be calculated by the content server as the users request contents from the content server.

## *C. Queueing model*

Let $V_i\left(t\right)$ be the random content arrival (number of bits) for user $i$ from the content server at the end of time slot $t$. We assume that each user can request at most one content during each time slot $t$ and, consequently, $V_i\left(t\right) \in \{0, L\}$. The queue length (i.e., number of bits) $Q_i\left(t\right)$ of user $i$ at the beginning of time slot $t$, will be given by [31]:

$$Q_i\left(t+1\right) = Q_i\left(t\right) - R_{ki}\left(t\right) + V_i\left(t\right), \tag{10}$$

where $R_{ki}\left(t\right)$ is the rate of user $i$. Since the content transmission links include: (a) UAV-user on the licensed band, (b) UAV-user on the unlicensed, (c) cloud-UAV-user on the unlicensed band, and (d) cloud-UAV-user on the licensed band as shown in Fig. 1, the rate of content transmission from UAV $k$ to user $i$ will be given by:

$$R_{ki}\left(u_{ki}\left(t\right), e_{ki}\left(t\right)\right) = \begin{cases} R_{lki}\left(u_{ki}\left(t\right)\right), & \text{link (a)}, \\ R_{uki}\left(e_{ki}\left(t\right)\right), & \text{link (b)}, \\ \frac{R_{uki}(e_{ki}(t)) R_{Ck}(t)}{R_{uki}(e_{ki}(t)) + R_{Ck}(t)}, & \text{link (c)}, \\ \frac{R_{lki}(u_{ki}(t)) R_{Ck}(t)}{R_{lki}(u_{ki}(t)) + R_{Ck}(t)}, & \text{link (d)}. \end{cases} \tag{11}$$

where the rate expression of link (c) is obtained from the fact that the time duration of a single data packet transmitted from the cloud to UAV $k$ is $1/R_{Ck}\left(t\right)$ and a single data packet transmitted from UAV $k$ to user $i$ is $1/R_{uki}\left(t\right)$. Therefore, the data rate of the transmission from the cloud to user $i$ is $\frac{1}{1/R_{Ck}(t) + 1/R_{uki}(t)}$.

From (11), we can see that the rate of any given user that receives a content from the UAV cache (link (a) and link (b)) is larger than the rate of another user that receives contents from the cloud (link (c) and link (d)). We use the notion of *queue stability* to measure the users' content transmission delay. In essence, a queue $Q_i\left(t\right)$ is said to be *rate stable* if [31]:

$$\lim_{t\to\infty}\frac{Q_i\left(t\right)}{t}=0. \tag{12}$$

From [31, Theorem 2.8], we can also see that the queue $Q_i\left(t\right)$ is rate stable if $R_{ki}\left(t\right)\geq V_i\left(t\right)$.

*D. Problem formulation*

Given this system model, our goal is to develop an effective spectrum allocation scheme for cache-enabled UAVs that can allocate appropriate bandwidth over the licensed and unlicensed bands to satisfy the queue stability requirement of each user. To achieve this goal, we formulate an optimization problem whose objective is to maximize the number of users that have stable queues. This maximization problem involves finding the optimal association $\mathcal{U}_k$ for each UAV $k$, bandwidth allocation indicators on the licensed band $\boldsymbol{u}_k$, time slot indicators on the unlicensed band $\boldsymbol{e}_k$, and the set of cached contents $\mathcal{C}_k$ for each UAV $k$. Therefore, this problem can be formalized as follows:

$$\max_{\boldsymbol{u}_k,\boldsymbol{e}_k,\mathcal{C}_k,\mathcal{U}_k}\sum_{k\in\mathcal{K}}\sum_{i\in\mathcal{U}_k}\mathbb{1}_{\left\{\lim\limits_{t\to\infty}\frac{Q_i(t)}{t}=0\right\}}=\sum_{k\in\mathcal{K}}\sum_{i\in\mathcal{U}_k}\mathbb{1}_{\{R_{ki}(u_{ki}(t),e_{ki}(t))\geq V_i(t)\}}, \tag{13}$$

$$\text{s. t.}\quad R_w\geq\gamma, \tag{13a}$$

$$\sum_{i\in\mathcal{U}_k}u_{ki}\left(t\right)\leq 1,\quad\forall k\in\mathcal{K}, \tag{13b}$$

$$\sum_{i\in\mathcal{U}_k}e_{ki}\left(t\right)\leq 1,\quad\forall k\in\mathcal{K}, \tag{13c}$$

where $\mathbb{1}_{\{x\}}=1$ when $x$ is true and $\mathbb{1}_{\{x\}}=0$ otherwise, $\mathcal{U}_k$ is the set of the users associated with UAV $k$, and $\boldsymbol{u}_k,\boldsymbol{e}_k$ denote the resource allocation indicators on the downlink licensed and unlicensed bands, respectively. (13a) guarantees that the average data rate of each WiFi user is above a desired threshold. In essence, LTE-U users can only occupy the unlicensed band when (13a) is satisfied. (13b) indicates that the licensed band allocation cannot exceed the total bandwidth for each UAV and (13c) captures the fact that the time slots over the unlicensed band cannot exceed the total number of time slots allocated to the UAVs.

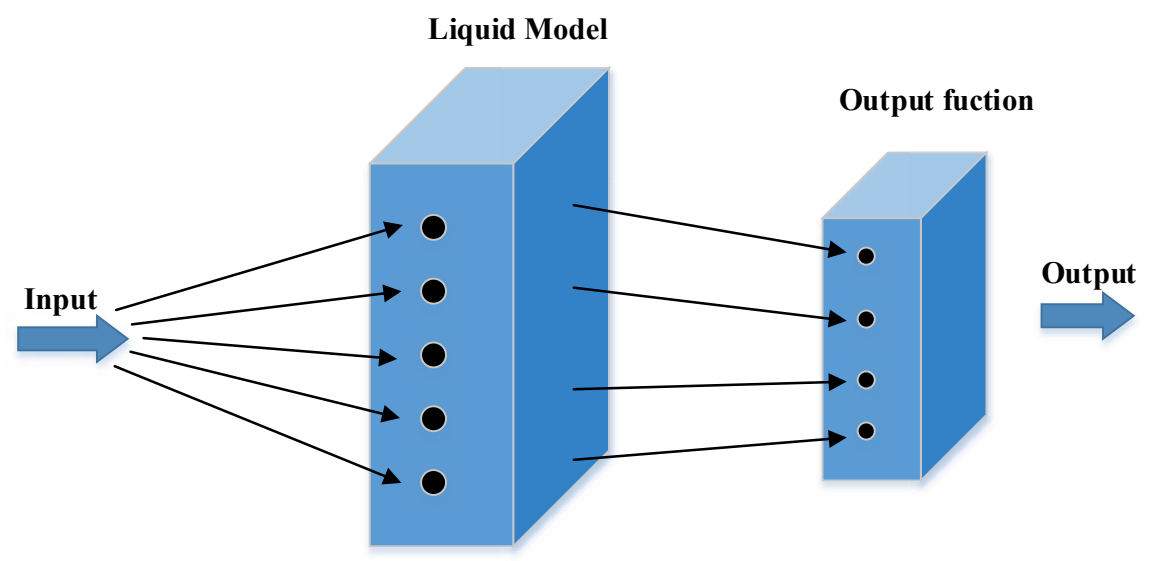

Fig. 3. The components of the proposed LSM-based algorithm.

## III. Liquid State Networks for Content prediction and Self-Organizing Resource Allocation

The optimization problem in (13) is challenging to solve, because spectrum allocation and content caching depend on the user association which, in turn, depends on the rate of each user. In fact, this problem can be shown to be combinatorial and non-convex, thus it is difficult to solve it using conventional optimization algorithms. Moreover, each UAV may not know the users' content requests which makes it challenging to determine which content to cache at the UAVs. To address these challenges, we propose a novel *liquid state machine learning* approach [32] to predict the users' content request distribution and perform resource allocation.

Liquid state machine is a novel kind of spiking neural networks [32] that are randomly generated. Learning algorithms based on LSM can store the users' behavioral information and track the state of a network over time. Therefore, an LSM-based algorithm will enable the cloud to leverage information on the users' behavior, that are stored in LSM, to predict the content request distribution and automatically adapt spectrum allocation to the any changes in the states of the network. Consequently, LSM-based algorithms are promising candidates for content request distribution prediction and wireless resource allocation problems.

Next, we first introduce the components of the proposed LSM algorithm. Then, we explain the entire process of using LSM to predict the users' content request distributions and to solve the problem in (13).

### A. LSM Components

As illustrated in Fig. 3, an LSM-based algorithm consists of five components: a) agents, b) input, c) output, d) liquid model, and e) output function. Since the prediction of the content request distribution and resource allocation are function-specific, we design the specific components for the problems of predicting content request distribution and resource allocation, separately.

*1) Content request distribution prediction:* The content request distribution prediction algorithm has the following components:

• *Agent*: In the studied system, the LSM agent is the cloud. Since each LSM approach performs a content request distribution prediction for just one user, the cloud must implement $U$ LSM algorithms.

• *Input:* The input of the LSM prediction algorithm for each user $i$ is defined by a vector $\boldsymbol{x}_i = [x_{i1}, \cdots, x_{iN_x}]^{\mathrm{T}}$ that captures the *context information* related to user $i$'s content request. Such information includes age, gender, occupation, and device type (e.g., tablet or smartphone). $N_x$ is the number of properties that constitute the context information of user $i$. The vector $\boldsymbol{x}_i$ is used to predict the content request distribution $\boldsymbol{y}_i$ of user $i$.

• *Output:* The LSM output is a vector of probabilities $\boldsymbol{y}_i = [\hat{p}_{i1}, \hat{p}_{i2}, \ldots, \hat{p}_{iN}]$ that represents the discrete probability density function of the content requests of user $i$ with $\hat{p}_{in}$ being the prediction of $p_{in}$ that is the probability of user $i$ requesting content $n$.

• *Liquid model:* A liquid model for each user $i$ can store the users' dynamic features that are extracted from the users' context over time. These dynamic features can be used with the output functions to predict the users' behavior such as content request and mobility patterns. Here, the liquid model consists of $W_1 \times W_2 \times W_3$ leaky integrate and fire neurons that are arranged in a 3D-column. In particular, each neuron that consists of resting state $S$, action state, and refractory period $T_f$ (ms) is defined as a excitatory neuron while the neuron that consists of resting state, inhibitory state, and refractory period is defined as an inhibitory neuron. The resting state indicates that the neuron does not receive any users' context information. The action state indicates that the neuron receives a certain amount of users' context and transmits this context to the connected neurons. The inhibitory state indicates that the users' information in the neuron is decreasing and, thus, cannot reach an action state. When each neuron transmits information to other connected neurons, its users' information may decrease to a value that is below that of the resting state. In this case, the neuron will have a refractory period during which the neuron returns to the resting state. The action state of neuron $j$ of LSM for user $i$, $v_{ij}(\tau)$ at time $\tau$ can

be given by [32]:

$$v_{ij}\left(\tau\right) = v_{ij}\left(\tau-1\right) + \frac{S + ZI_i\left(\tau-1\right) - v_{ij}\left(\tau-1\right)}{Z\rho}, \tag{14}$$

where $Z$ is the neuron resistance, $I_i\left(\tau-1\right)$ is the input of the users' information, and $\rho$ is the neuron time constant. Based on (14), the LSM state can be given by $\boldsymbol{v}_i(\tau) = \left[v_{i1}\left(\tau\right), \ldots, v_{iN_W}(\tau)\right]$, where $N_W = W_1 \times W_2 \times W_3$ is the number of neurons. Note that, the time duration of $\tau$ is much smaller than the time duration of $t$.

The connections from the input to the liquid model are made with probability $\mathbb{P}_{\text{IN}}$. The probability connection between neurons $i$ and $j$ can be given by:

$$\mathbb{P}_{ij} = \varsigma e^{-\left(d(i,j)/\lambda\right)^2}, \tag{15}$$

where $\varsigma \in \left\{\varsigma_{\text{EE}}, \varsigma_{\text{EI}}, \varsigma_{\text{IE}}, \varsigma_{\text{II}}\right\}$ is a constant that depends on the type of both neurons. In particular, $\varsigma_{\text{EE}}$ denotes an excitatory-excitatory connection, $\varsigma_{\text{EI}}$ is an excitatory-inhibitory connection, $\varsigma_{\text{IE}}$ is an inhibitory-excitatory connection, and $\varsigma_{\text{II}}$ is a inhibitory-inhibitory connection. $d\left(i,j\right)$ is the Euclidean distance between neurons $i$ and $j$. $\lambda$ is a parameter that influences how often neurons are connected.

• *Output function:* The output function is used to build the relationship between the state of the LSM model and the prediction of each user's content request distribution. Let $\boldsymbol{f}_i \in \mathbb{R}^{N_W \times N}$ be the output function of LSM, where $N$ is the total number of contents. In order to predict the user's content request distribution, $\boldsymbol{f}_i$ is trained in an offline manner using ridge regression to approximate the prediction function:

$$\boldsymbol{f}_i = \boldsymbol{y}_{T,i}\boldsymbol{v}_i^{\mathrm{T}}\left(\boldsymbol{v}_i^{\mathrm{T}}\boldsymbol{v}_i + \delta^2\boldsymbol{I}\right)^{-1}, \tag{16}$$

where $\boldsymbol{v}_i = \left[\boldsymbol{v}_i\left(1\right), \ldots, \boldsymbol{v}_i\left(N_t\right)\right]$ is the LSM state sequence for user $i$. Here, $\boldsymbol{y}_{T,i}$ is the target output of the LSM algorithm, $\boldsymbol{I}$ is an identity matrix, and $\delta$ is the learning rate. Based on the trained output function, the prediction of user $i$'s content request distribution can be given by:

$$\boldsymbol{y}_i = \boldsymbol{f}_i\boldsymbol{v}_i. \tag{17}$$

Given the above components, the LSM based algorithm first uses (14) to store the users' context information and, then, uses (16) to train the output function of the LSM. After training, the LSM algorithm can predict the users' content request distribution based on the users' context information.

*2) Resource allocation:* The LSM reinforcement learning algorithm for resource allocation has the following components:

• *Agent*: The agents in the LSM algorithm used for resource allocation are the UAVs.

• *Input:* The LSM input is $\boldsymbol{m}_k(t)=[\boldsymbol{\pi}_1(t),\cdots,\boldsymbol{\pi}_{k-1}(t),\boldsymbol{\pi}_{k+1}(t),\cdots,\boldsymbol{\pi}_K(t),V_1(t),\cdots,V_{U_k}(t)]^{\mathrm{T}}$ where $\boldsymbol{\pi}_k(t)=\left[\pi_{k,a_{k1}}(t),\ldots,\pi_{k,a_{kA_k}}(t)\right]$ is a probability distribution for UAV $k$. If action $a_{kj}$ can maximize the average number of stable queue users, then $\pi_{k,a_{kj}}=1-\varepsilon+\frac{\varepsilon}{A_k}$, $\pi_{k,a_{kj}}=\frac{\varepsilon}{A_k}$, otherwise. $A_k$ is the number of actions that each UAV $k$ can take and $U_k$ is the number of the users that can associate with UAV $k$ which will be specified in Theorem 3. In particular, each UAV's action represents a UAV-user association scheme.

• *Output:* The LSM output at time $t$ is a vector $\boldsymbol{b}_k(t)=[b_{k1}(t),b_{k2}(t),\ldots,b_{kA_k}(t)]$ that represents the resource allocation results. $b_{kj}(t)$ is the expected number of stable queue users when UAV $k$ uses user association scheme $j$, which is:

$$b_{kj}(t)=\sum_{\boldsymbol{a}_{-k}\in\mathcal{A}_{-k}}b_{kj,\boldsymbol{a}_{-k}}(a_{kj},\boldsymbol{a}_{-k})\,\pi_{-k,\boldsymbol{a}_{-k}}(t), \tag{18}$$

where $\mathcal{A}_{-k}$ is the action set of all UAVs other than UAV $k$ and $b_{kj,\boldsymbol{a}_{-k}}$ is the number of stable queue users as UAV $k$ uses action $a_{kj}$ and the other UAVs use actions $\boldsymbol{a}_{-k}$. $\pi_{-k,\boldsymbol{a}_{-k}}(t)=\prod\limits_{n\in\mathcal{K},n\neq k,}\pi_{n,a_n}(t)$ is the marginal probability distribution over the action set of UAV $k$. To calculate $b_{kj,\boldsymbol{a}_{-k}}(a_{kj},\boldsymbol{a}_{-k})$, UAV $k$ needs to determine the optimal contents to cache as well as the resource allocation for any given user association $a_{kj}$. Based on the content request distribution $\boldsymbol{y}_i$ obtained by the LSM-based prediction algorithm mentioned before, the optimal contents stored at UAV cache for a given user association can be determined by the following theorem:

**Theorem 1.** Given $\mathcal{U}_k$, the user association of each UAV $k$, $\boldsymbol{y}_i$, the content request distribution of user $i$, and $N_i^{\mathrm{C}}$, the number of contents that user $i$ requests, the optimal contents stored at each UAV $k$ can be given by:

$$\mathcal{C}_k=\arg\max_{\mathcal{C}_k}\sum_{n\in\mathcal{C},i\in\mathcal{U}_k}\left(\hat{p}_{in}N_i^C\right), \tag{19}$$

where $\mathcal{C}_k$ is the set of contents that are stored at the cache of UAV $k$.

*Proof.* The contents stored at the cache are used to reduce the content transmission over fronthaul links from the cloud to UAV $k$. From (9), we can see that the bandwidth of the fronthaul is equally allocated to those users that request contents from the cloud and, hence, the objective of the optimal content caching policy is to minimize the number of the users that request contents from the cloud. For each user $i$, the average number of the requested contents is $\boldsymbol{y}_i N_i^{\mathrm{C}} = \left[\hat{p}_{i1} N_i^{\mathrm{C}}, \ldots, \hat{p}_{iN} N_i^{\mathrm{C}}\right]$ where $\hat{p}_{in} N_i^{\mathrm{C}}$ is the average number of requests of content $n$. Therefore, the total average number of content requests for the users associated with UAV $k$ is $\sum\limits_{i \in \mathcal{U}_k} \left(\boldsymbol{y}_i N_i^{\mathrm{C}}\right) = \left[\sum\limits_{i \in \mathcal{U}_k} \left(\hat{p}_{i1} N_i^{\mathrm{C}}\right), \ldots, \sum\limits_{i \in \mathcal{U}_k} \left(\hat{p}_{iN} N_i^{\mathrm{C}}\right)\right]$. Clearly, UAV $k$ will choose the contents that have the highest average number of requests to store at cache which can be given by $\max\limits_{\mathcal{C}_k} \sum\limits_{n \in \mathcal{C}, i \in \mathcal{U}_k} \left(\hat{p}_{in} N_i^{\mathrm{C}}\right)$. This completes the proof. □

From Theorem 1, we can see that the optimal caching policy depends only on the users' association policies and their content request distributions. When the user association and cached contents [33] are determined, the transmission link for each content (i.e., from cloud or UAV cache) is determined. $\mathcal{U}_{kc}$ ($\mathcal{U}_{kk}$) is defined as the set of users whose requested contents are (not) stored at cache of UAV $k$. Given a determined user association $a_{kj}$, (13) for each UAV $k$ can be simplified as:

$$\max_{\boldsymbol{u}_k, \boldsymbol{e}_k} \sum_{i \in \mathcal{U}_{kc}} \mathbb{1}_{\left\{R_{lki}(u_{ki}(t)) = V_i(t) \text{ or } R_{uki}(e_{ki}(t)) = V_i(t)\right\}} + \sum_{i \in \mathcal{U}_{kk}} \mathbb{1}_{\left\{\frac{R_{lki}(u_{ki}(t)) R_{Ck}(t)}{R_{lki}(u_{ki}(t)) + R_{Ck}(t)} = V_i(t) \text{ or } \frac{R_{uki}(e_{ki}(t)) R_{Ck}(t)}{R_{uki}(e_{ki}(t)) + R_{Ck}(t)} = V_i(t)\right\}}. \tag{20}$$

The first term in (20) indicates that the users' requested contents are transmitted from the cache of the UAV. The second term indicates that the requested contents are transmitted from the cloud.

To solve the problem in (20), we propose a search-based algorithm. Then, we prove that the solution of the proposed algorithm is an optimal solution for (20). We define the vector $\boldsymbol{u}_k^{\mathrm{R}}(t) = \left[u_{k1}^{\mathrm{R}}(t), \ldots, u_{kU_k}^{\mathrm{R}}(t)\right]$ with each element being the fraction of the licensed spectrum that UAV $k$ uses to guarantee stable queues for its associated users where $u_{ki}^{\mathrm{R}}(t) = \frac{V_i(t)}{R_{lki}(1)}$ is the licensed spectrum that UAV $k$ uses to guarantee a stable queue for user $i$. Similarly, $\boldsymbol{e}_k^{\mathrm{R}}(t) = \left[e_{k1}^{\mathrm{R}}(t), \ldots, e_{kU_k}^{\mathrm{R}}(t)\right]$ is a vector with each element being the fraction of the unlicensed spectrum that UAV $k$ uses to guarantee stable queues for the users where $e_{ki}^{\mathrm{R}}(t) = \frac{V_i(t)}{R_{uki}(1)}$ is the fraction of the unlicensed spectrum that UAV $k$ uses to guarantee a stable queue for user $i$. Then, $\boldsymbol{u}_k^{\mathrm{MR}}(t) = \left[u_{k1}^{\mathrm{MR}}(t), \ldots, u_{kU_k}^{\mathrm{MR}}(t)\right]$ is defined as a vector with each element $u_{ki}^{\mathrm{MR}}(t)$ being the minimum licensed bandwidth that UAV $k$ needs to guarantee a stable queue for user $i$ where

$$u_{ki}^{\mathrm{MR}}(t) = \begin{cases} u_{ki}^{R}(t), u_{ki}^{R}(t) \leq e_{ki}^{R}(t), \\ 0, \qquad \text{else.} \end{cases} \tag{21}$$

From (21), we can see that, as the licensed spectrum that UAV $k$ uses to guarantee a stable queue for user $i$ is smaller than the unlicensed spectrum that UAV $k$ needs to guarantee a stable queue for user $i$ ($u_{ki}^{R}(t) < e_{ki}^{R}(t)$), $u_{ki}^{\mathrm{MR}}(t) = u_{ki}^{R}(t)$. Similarly, $\boldsymbol{e}_{k}^{\mathrm{MR}}(t) = \left[e_{k1}^{\mathrm{MR}}(t), \ldots, e_{kU_k}^{\mathrm{MR}}(t)\right]$ is defined as a vector with each element $e_{ki}^{\mathrm{MR}}(t)$ being the minimum unlicensed bandwidth that UAV $k$ uses to guarantee a stable queue for user $i$ with

$$e_{ki}^{\mathrm{MR}}(t) = \begin{cases} e_{ki}^{R}(t), e_{ki}^{R}(t) < u_{ki}^{R}(t), \\ 0, \qquad \text{else}. \end{cases} \tag{22}$$

We also define $\boldsymbol{e}_{k,\max}^{\mathrm{MR}}(t) = \arg\max\limits_{\boldsymbol{e}_{k,\max}^{\mathrm{MR}}(t)} \sum\limits_{i=1}^{U_k} \mathbb{1}_{\left\{e_{ki,\max}^{\mathrm{MR}}(t)>0\right\}}$ with $\sum\limits_{i=1}^{U_k} e_{ki,\max}^{\mathrm{MR}}(t) \le 1$ and $e_{ki,\max}^{\mathrm{MR}}(t) \in \{0, e_{ki}^{\mathrm{MR}}(t)\}$ where $\mathbb{1}_{\{x\}} = 1$ when $x$ is true and $\mathbb{1}_{\{x\}} = 0$, otherwise. $e_{ki,\max}^{\mathrm{MR}}(t)$ is the element $i$ of $\boldsymbol{e}_{k,\max}^{\mathrm{MR}}(t)$. The definition of $\boldsymbol{u}_{k,\max}^{\mathrm{MR}}(t)$ is similar to $\boldsymbol{e}_{k,\max}^{\mathrm{MR}}(t)$. $N_{\boldsymbol{e}_{k,\max}^{\mathrm{MR}}(t)}$ and $N_{\boldsymbol{u}_{k,\max}^{\mathrm{MR}}(t)}$ are defined as the number of vectors $\boldsymbol{u}_{k,\max}^{\mathrm{MR}}(t)$ and $\boldsymbol{e}_{k,\max}^{\mathrm{MR}}(t)$, respectively. $N_{\boldsymbol{e}_{k,\max}^{\mathrm{MR}}(t)}$ and $N_{\boldsymbol{u}_{k,\max}^{\mathrm{MR}}(t)}$ will directly determine the number of searches needed by the search-based algorithm. We define $\tilde{\boldsymbol{u}}_{k}^{\mathrm{R}}\left(\boldsymbol{e}_{k,\max}^{\mathrm{MR}}(t)\right) = \left[\tilde{u}_{k1}^{\mathrm{R}}\left(e_{k1,\max}^{\mathrm{MR}}(t)\right), \ldots, \tilde{u}_{kU_k}^{\mathrm{R}}\left(e_{kU_k,\max}^{\mathrm{MR}}(t)\right)\right]$ with

$$\tilde{u}_{ki}^{\mathrm{R}}\left(e_{ki,\max}^{\mathrm{MR}}(t)\right) = \begin{cases} u_{ki}^{R}(t), u_{ki,\max}^{\mathrm{MR}}(t) = 0, \\ 0, \qquad \text{else}. \end{cases} \tag{23}$$

Given $\tilde{\boldsymbol{u}}_{k}^{\mathrm{R}}\left(\boldsymbol{e}_{k,\max}^{\mathrm{MR}}(t)\right)$, we define $\tilde{\boldsymbol{u}}_{k,\max}^{\mathrm{R}}\left(\boldsymbol{e}_{k,\max}^{\mathrm{MR}}(t)\right) = \arg\max\limits_{\tilde{\boldsymbol{u}}_{k,\max}^{\mathrm{R}}\left(\boldsymbol{e}_{k,\max}^{\mathrm{MR}}(t)\right)} \sum\limits_{i=1}^{U_k} \mathbb{1}_{\left\{\tilde{u}_{ki,\max}^{\mathrm{R}}\left(\boldsymbol{e}_{k,\max}^{\mathrm{MR}}(t)\right)>0\right\}}$ with $\sum\limits_{i=1}^{U_k} \tilde{u}_{ki,\max}^{\mathrm{R}}\left(\boldsymbol{e}_{k,\max}^{\mathrm{MR}}(t)\right) \le 1$ and $\tilde{u}_{ki,\max}^{\mathrm{R}}\left(\boldsymbol{e}_{k,\max}^{\mathrm{MR}}(t)\right) \in \left\{0, \tilde{u}_{ki}^{\mathrm{R}}\left(\boldsymbol{e}_{k,\max}^{\mathrm{MR}}(t)\right)\right\}$. Based on these defined vectors, the search-based algorithm is given in Algorithm 1. The solution of Algorithm 1 is an optimal solution of the problem in (20) and can be characterized by the following theorem:

**Theorem 2.** Given the vectors $\boldsymbol{u}_{k}^{\mathrm{R}}(t)$, $\boldsymbol{u}_{k}^{\mathrm{MR}}(t)$, $\boldsymbol{u}_{k,\max}^{\mathrm{MR}}(t)$ $\boldsymbol{e}_{k}^{\mathrm{R}}(t)$, $\boldsymbol{e}_{k}^{\mathrm{MR}}(t)$, and $\boldsymbol{e}_{k,\max}^{\mathrm{MR}}(t)$ as well as $N_{\boldsymbol{u}_{k,\max}^{\mathrm{MR}}(t)}$ and $N_{\boldsymbol{e}_{k,\max}^{\mathrm{MR}}(t)}$, the optimal resource allocation vectors for UAV $k$, $\boldsymbol{u}^{*}(t)$ and $\boldsymbol{e}^{*}(t)$ can be given by:

i) If $\sum\limits_{i \in U_k} u_{ki}^{\mathrm{MR}}(t) \le 1$ and $\sum\limits_{i \in U_k} e_{ki}^{\mathrm{MR}}(t) \le 1$, then $\boldsymbol{u}_{k}^{*}(t) = \boldsymbol{u}_{k}^{\mathrm{MR}}(t)$ and $\boldsymbol{e}_{k}^{*}(t) = \boldsymbol{e}_{k}^{\mathrm{MR}}(t)$.

ii) If $N_{\boldsymbol{u}_{k,\max}^{\mathrm{MR}}(t)} > 1$, then $[\boldsymbol{u}_{k}^{*}, \boldsymbol{e}_{k}^{*}] = \arg\max\limits_{\boldsymbol{u}_{k,\max}^{\mathrm{MR}}(t), \tilde{\boldsymbol{e}}_{k}^{\mathrm{R}}\left(\boldsymbol{u}_{k,\max}^{\mathrm{MR}}(t)\right)} \left(n\left(\boldsymbol{u}_{k,\max}^{\mathrm{MR}}(t)\right) + n\left(\tilde{\boldsymbol{e}}_{k}^{\mathrm{R}}\left(\boldsymbol{u}_{k,\max}^{\mathrm{MR}}(t)\right)\right)\right)$, where $n(\boldsymbol{x})$ represents the number of elements in the support of vector $\boldsymbol{x}$.

iii) If $N_{\boldsymbol{e}_{k,\max}^{\mathrm{MR}}(t)} > 1$, then $[\boldsymbol{e}_{k}^{*}, \boldsymbol{u}_{k}^{*}] = \arg\max\limits_{\boldsymbol{e}_{k,\max}^{\mathrm{MR}}(t), \tilde{\boldsymbol{u}}_{k}^{\mathrm{R}}\left(\boldsymbol{e}_{k,\max}^{\mathrm{MR}}(t)\right)} \left(n\left(\boldsymbol{e}_{k,\max}^{\mathrm{MR}}(t)\right) + n\left(\tilde{\boldsymbol{u}}_{k}^{\mathrm{R}}\left(\boldsymbol{e}_{k,\max}^{\mathrm{MR}}(t)\right)\right)\right)$.

**Algorithm 1** Search-based algorithm

**Initialization:** $\boldsymbol{u}_k^{\rm R}(t)$, $\boldsymbol{e}_k^{\rm R}(t)$, $\boldsymbol{u}_k^{\rm MR}(t)=0$, $\boldsymbol{e}_k^{\rm MR}(t)=0$.

Set $\boldsymbol{u}_k^{\rm MR}(t)$ and $\boldsymbol{e}_k^{\rm MR}(t)$ based on (21) and (22).
Set $\boldsymbol{u}_{k,\max}^{\rm MR}(t)$, $\boldsymbol{e}_{k,\max}^{\rm MR}(t)$, $N_{\boldsymbol{u}_{k,\max}^{\rm MR}(t)}$, and $N_{\boldsymbol{e}_{k,\max}^{\rm MR}(t)}$ .
**if** $\sum\limits_{i\in U_k} u_{ki}^{\rm MR}(t)\le 1$ and $\sum\limits_{i\in U_k} e_{ki}^{\rm MR}\le 1$ **then**
    $\boldsymbol{u}_k^{*}=\boldsymbol{u}_k^{\rm MR}$, $\boldsymbol{e}_k^{*}=\boldsymbol{e}_k^{\rm MR}$.
**else if** $N_{\boldsymbol{u}_{k,\max}^{\rm MR}(t)}>1$ **then**
    Traverse all $\boldsymbol{u}_{k,\max}^{\rm MR}(t)$ and modify $\boldsymbol{e}_{k,\max}^{\rm MR}(t)$ to find the maximum nonzero elements in $\boldsymbol{u}_{k,\max}^{\rm MR}(t)$ and $\boldsymbol{e}_{k,\max}^{\rm MR}(t)$.
**else if** $N_{\boldsymbol{e}_{k,\max}^{\rm R'}(t)}>1$ **then**
    Traverse all $\boldsymbol{e}_{k,\max}^{\rm MR}(t)$ and modify $\boldsymbol{u}_{k,\max}^{\rm MR}(t)$ to find the maximum nonzero elements in $\boldsymbol{u}_{k,\max}^{\rm MR}(t)$ and $\boldsymbol{e}_{k,\max}^{\rm MR}(t)$.
**end if**

*Proof.* For case i), If $\sum\limits_{i\in U_k} u_{ki}^{\rm MR}(t)\le 1$ and $\sum\limits_{i\in U_k} e_{ki}^{\rm MR}(t)\le 1$, then UAV $k$ can guarantee stable queues for all of its users. Therefore, $\boldsymbol{u}_k^{*}(t)=\boldsymbol{u}_k^{\rm MR}(t)$, $\boldsymbol{e}_k^{*}(t)=\boldsymbol{e}_k^{\rm MR}(t)$. For case ii), $N_{\boldsymbol{u}_{k,\max}^{\rm MR}(t)}>1$ indicates that the optimal resource allocation vector $\boldsymbol{u}_{k,\max}^{\rm MR}(t)$ is not unique. Therefore, UAV $k$ needs to traverse all of the vectors $\boldsymbol{u}_{k,\max}^{\rm MR}(t)$ and its corresponding vector $\tilde{\boldsymbol{e}}_k^{\rm R}\left(\boldsymbol{u}_{k,\max}^{\rm MR}(t)\right)$. In $\boldsymbol{u}_{k,\max}^{\rm MR}(t)$, as shown in (21), $u_{ki}^{\rm R}(t)<e_{ki}^{\rm R}(t)$. Hence, when UAV $k$ changes the band allocated to a user from the licensed band to the unlicensed band, it will use more unlicensed resources than licensed resources to guarantee a stable queue for this user. In consequence, when $n\left(\boldsymbol{u}_{k,\max}^{\rm MR'}(t)\right)=n\left(\boldsymbol{u}_{k,\max}^{\rm MR}(t)\right)-1$, we have $n\left(\tilde{\boldsymbol{e}}_k^{\rm R}\left(\boldsymbol{u}_{k,\max}^{\rm MR'}(t)\right)\right)\le n\left(\tilde{\boldsymbol{e}}_k^{\rm R}\left(\boldsymbol{u}_{k,\max}^{\rm MR}(t)\right)\right)+1$ where $\boldsymbol{u}_{k,\max}^{\rm MR'}(t)$ and $\boldsymbol{u}_{k,\max}^{\rm MR}(t)$ differ by only one element that pertains to the user whose spectrum allocation has changed. Then, we can obtain that $n\left(\boldsymbol{u}_{k,\max}^{\rm MR'}(t)\right)+n\left(\tilde{\boldsymbol{e}}_k^{\rm R}\left(\boldsymbol{u}_{k,\max}^{\rm MR'}(t)\right)\right)\le n\left(\boldsymbol{u}_{k,\max}^{\rm MR}(t)\right)+n\left(\tilde{\boldsymbol{e}}_k^{\rm R}\left(\boldsymbol{u}_{k,\max}^{\rm MR}(t)\right)\right)$. Therefore, $\arg\max\limits_{\boldsymbol{u}_{k,\max}^{\rm MR,}(t),\tilde{\boldsymbol{e}}_k^{\rm R}\left(\boldsymbol{u}_{k,\max}^{\rm MR}(t)\right)}\left(n\left(\boldsymbol{u}_{k,\max}^{\rm MR}(t)\right)+n\left(\tilde{\boldsymbol{e}}_k^{\rm R}\left(\boldsymbol{u}_{k,\max}^{\rm MR}(t)\right)\right)\right)$ is the optimal resource allocation for UAV $k$. Case iii) can be proved in a similar way to case ii). This completes the proof. □

From Theorem 2, we can see that the complexity of the search-based algorithm depends on the number of vectors $\boldsymbol{u}_{k,\max}^{\rm MR}(t)$ or $\boldsymbol{e}_{k,\max}^{\rm MR}(t)$. Moreover, the resource allocation over the licensed band $\boldsymbol{u}_{k,\max}^{{\rm R}M}(t)$ is related to the resource allocation via the unlicensed band $\tilde{\boldsymbol{e}}_k^{\rm R}\left(\boldsymbol{u}_{k,\max}^{\rm MR}(t)\right)$. Note that, the optimal resource allocation for the problem in (20) is not unique. Theorem 2 provides a method to find an optimal resource allocation for a given user association but not to find all of the optimal resource allocations for a given user association. Based on Theorem 1 and the search-based algorithm, the optimal caching contents and resource allocation for a given user association $a_{kj}$ are determined and, hence, $b_{kj,\boldsymbol{a}_{-k}}\left(a_{kj},\boldsymbol{a}_{-k}\right)$ in (18) can be calculated. Given $b_{kj,\boldsymbol{a}_{-k}}\left(a_{kj},\boldsymbol{a}_{-k}\right)$, the output of the LSM resource allocation algorithm can be obtained. Given the input and output, the LSM algorithm can find the relationship between them using the liquid model and output function that are introduced next.

- *Liquid model:* The liquid model in the resource allocation algorithm is used to store the network state information including UAV-user association schemes and their corresponding output results. The liquid model consists of $W_1^\alpha \times W_2^\alpha \times W_3^\alpha$ number of leaky integrate and fire neurons that are arranged in a 3D-column. The generation of the resource allocation LSM model is similar to the one in the content request distribution prediction case.

- *Output function:* The output function is used to build the relationship between the UAVs-user association schemes and the number of users with a stable queue. Let $\boldsymbol{f}_k^\alpha \in \mathbb{R}^{A_k \times \left(A_k + N_W^\alpha\right)}$ be the output function of UAV $k$, where $N_W^\alpha$ is the number of the neurons in the liquid model. To train $\boldsymbol{f}_k^\alpha$, a linear gradient descent approach can be used to derive the following update rule,

$$\boldsymbol{f}_{k,i}^\alpha\left(t+1\right) = \boldsymbol{f}_{k,i}^\alpha\left(t\right) + \delta^\alpha\left(e_{k,i}\left(t\right) - b_{ki}\left(t\right)\right)\left[\boldsymbol{v}_k\left(t\right);\boldsymbol{m}_k\left(t\right)\right]^{\mathrm{T}}, \tag{24}$$

where $\boldsymbol{f}_{k,i}^\alpha$ is row $i$ of $\boldsymbol{f}_k^\alpha$, $\delta^\alpha$ is the learning rate, $e_{k,i}\left(t\right)$ is the expected output, and $\boldsymbol{v}_k\left(t\right)$ is the LSM state at time $t$. From (24), we can see that the construction of $\boldsymbol{f}_k^\alpha$ depends on the number of actions of UAV $k$. Therefore, we first need to determine the number of actions for each UAV $k$ which can be given in the following theorem:

**Theorem 3.** Given the locations of UAVs and users, the number of actions for UAV $k$ can be given by:

$$A_k = 2^{\left|\mathcal{U}_k\right|}, \tag{25}$$

where $\mathcal{U}_k = \begin{cases} \left\{i \left| R_{lki}\left(1\right) \geq L \text{ or } R_{uki}\left(1\right) \geq L \right.\right\}, & R_{Ck,\max} \gg R_{lki}\left(1\right) \text{ and } R_{Ck,\max} \gg R_{uki}\left(1\right), \\ \left\{i \left| \frac{R_{lki}(1)R_{Ck,\max}}{R_{lki}(1)+R_{Ck,\max}} \geq L \text{ or } \frac{R_{uki}(1)R_{Ck,\max}}{R_{uki}(1)+R_{Ck,\max}} \geq L \right.\right\}, & \text{else,} \end{cases}$

$R_{Ck,\max} = F_C \log_2\left(1 + \frac{P_C \bar{L}_k}{\sum\limits_{j\in\mathcal{K}, j\neq k} P_K 10^{\bar{l}_{ki}^{l}/10} + \sigma^2}\right)$, and $\left|\mathcal{U}_k\right|$ is the number of elements in $\mathcal{U}_k$.

*Proof.* To calculate the number of actions of UAV $k$, we need to first determine the set of the users that can associate with UAV $k$. From (13), we can see that, for user $i$ associated with UAV $k$, the maximum data rate $R_{ki,\max}$ in (11) must be larger than or equal to $L$. Here, we only need to consider the maximum data rate $R_{ki,\max}$ of links (c) and (d). This is due to the fact that when the data rate of links (c) or (d) meets the data rate requirement, the data rate of links (a) or (b) will also meet the data rate requirement. Hence, $R_{ki,\max} = \begin{cases} \frac{R_{uki}(1)R_{Ck,\max}}{R_{uki}(1)+R_{Ck,\max}}, & \text{link (c)}, \\ \frac{R_{lki}(1)R_{Ck,\max}}{R_{lki}(1)+R_{Ck,\max}}, & \text{link (d)}. \end{cases}$ where $R_{lki}\left(1\right)$ and $R_{uki}\left(1\right)$ imply that all of the bandwidth of the licensed and unlicensed bands is allocated to user $i$ (i.e., $u_{ki}\left(t\right)=1$ and $e_{ki}\left(t\right)=1$) and $R_{Ck,\max}=F_C\log_2\left(1+\frac{P_C\bar{L}_k}{\sum\limits_{j\in\mathcal{K},j\neq k} P_K 10^{\bar{l}^l_{ki}/10}+\sigma^2}\right)$ indicates that all of the fronthaul bandwidth $F_C$ is allocated to user $i$. User $i$ can associate with UAV $k$ when the data rate $R_{Ck,\max}$ of links (c) or (d) meets the rate requirement. Therefore, we have $\mathcal{U}_k=\left\{i\left|\frac{R_{lki}(1)R_{Ck,\max}}{R_{lki}(1)+R_{Ck,\max}}\geq L \text{ or } \frac{R_{uki}(1)R_{Ck,\max}}{R_{uki}(1)+R_{Ck,\max}}\geq L\right.\right\}$. As $R_{Ck,\max}\gg R_{lki}\left(1\right)$, $\frac{R_{lki}(1)R_{Ck,\max}}{R_{lki}(1)+R_{Ck,\max}}$ can be simplified as $\frac{R_{lki}(1)R_{Ck,\max}}{R_{Ck,\max}}=R_{lki}\left(1\right)$. Similarly, $R_{Ck,\max}\gg R_{uki}\left(1\right)$, $\frac{R_{uki}(1)R_{Ck,\max}}{R_{uki}(1)+R_{Ck,\max}}$ can be simplified as $R_{uki}\left(1\right)$. Hence, when $R_{Ck,\max}\gg R_{uki}\left(1\right),R_{lki}\left(1\right)$, $\mathcal{U}_k=\left\{i\left|R_{lki}\left(1\right)\geq L \text{ or } R_{uki}\left(1\right)\geq L\right.\right\}$. Each user $i\in\mathcal{U}_k$ can associate with UAV $k$ or associate with other UAVs, and, hence, total number of actions (user association schemes) is $2^{|\mathcal{U}_k|}$ where $|\mathcal{U}_k|$ represents the number of elements (users) in $\mathcal{U}_k$. This completes the proof. $\square$

From Theorem 3, we can see that, when the data rate of fronthaul $R_{Ck}$ is much larger than the data rate of the links from UAVs to users ($R_{lk}$ or $R_{uk}$), the number of actions for each UAV $k$ only depends on $R_{lk}$ or $R_{uk}$. In particular, the bandwidths of fronthaul, licensed, and unlicensed bands will significantly affect the number of the users that can associate with UAV $k$. Based on Theorem 3, the output function $\boldsymbol{f}_k^{\alpha}$ can be trained using (24).

After the training of the output function $\boldsymbol{f}_k^{\alpha}$, the estimated number of users that have a stable queue given user association $a_{kj}$ is:

$$b_{kj}\left(t\right)=\boldsymbol{f}_{k,j}^{\alpha}\left[\boldsymbol{v}_k\left(t\right);\boldsymbol{m}_k\left(t\right)\right]. \tag{26}$$

### B. LSM Algorithm for content prediction and spectrum allocation

Based on the components of the proposed LSM algorithms detailed in Section III-A, next, we explain the entire process of using LSM algorithms to solve the problem in (13). To solve the problem in (13), the cloud first predicts the content request distribution of each user using the proposed LSM-based prediction approach. Based on the users' content request distribution, each UAV $k$ uses the LSM-based learning algorithm with $\epsilon$-greedy mechanism [23] to find the optimal user association policy. At each iteration, each UAV $k$ takes one action (users association) $a_{kj}$ based on the $\epsilon$-greedy mechanism. Once the users association is determined, the

**Algorithm 2** LSM-based learning algorithm

**Input:** The set of users' context, $\boldsymbol{x}_j(t)$, UAVs' input $\boldsymbol{m}_k(t)$.

**Initialization:** The cloud generates the liquid model for each user.

Each UAV generates a liquid model based on (14) and (15).

1: Calculate the time slots $L$ based on (2).
2: Predict users' content request distribution using (17).
3: **for** time $t$ **do**
4: Estimate the number of the users that are at stable queuing state using (26).
5: **if** $rand(.) < \varepsilon$ **then**
6: Randomly choose one action.
7: **else**
8: Choose action $a_k(t) = \underset{a_k(t)}{\arg\max}(\boldsymbol{b}_k(t))$.
9: **end if**
10: Calculate the number of the stable queue users based on Theorem 1 and search-based algorithm.
11: Update the output weight matrix $\boldsymbol{f}^{\alpha}_{k,i}(t)$ based on (24).
12: Update the input $\boldsymbol{a}_k(t)$ according to the result of the users choosing UAVs.
13: **end for**

optimal contents stored at the cache of UAV $k$ can be determined by Theorem 1 and the optimal spectrum allocation will also be determined by the search-based algorithm. Based on the user association, optimal resource allocation and optimal caching contents, each UAV $k$ can calculate the average number of stable queue users $b_{kj,\boldsymbol{a}_{-k}}(a_{kj}, \boldsymbol{a}_{-k})$ when UAV $k$ uses action $a_{kj}$. In this algorithm, each UAV can store the users' and network's states as UAV $k$ adopts different users association schemes. During each iteration, the LSM algorithm can record the number of stable queue users, $b_{kj,\boldsymbol{a}_{-k}}(a_{kj}, \boldsymbol{a}_{-k})$. Since the LSM-based algorithm satisfies the convergence conditions of [23, Theorem 2], as time elapses, each UAV $k$'s output resulting from the resource allocation scheme $j$ will converge to a final value $b_{kj}$. At this convergence point, $b_{kj}$ indicates the expected value of the number of stable queue users with respect to all other UAVs' strategies. The proposed LSM approach performed by the cloud and each UAV $k$ is shown, in detail, in Algorithm 1. In line 5, $rand(.)$ represents a number chosen randomly from $[0, 1]$.

## IV. Simulation Results

In our simulations, the content request data that the LSM uses to train and predict content request distribution is obtained from *Youku* of *China network video index*[1]. We consider a circular cloud-based UAVs network area having a radius $r = 200$ m, $U = 20$ uniformly distributed users and $K = 5$ uniformly distributed UAVs. For implementing the proposed LSM-based algorithm, we use the MATLAB LSM toolbox described in [32]. Other system parameters are listed in Table II. For comparison purpose, we use two baselines:

- The first baseline is Q-learning algorithm in [34] with content caching, which we refer to as "Q-learning with cache". This Q-learning algorithm is used to find the optimal resource allocation and user association for each UAV. For this Q-learning algorithm, the state is set to the LSM's input $\boldsymbol{m}_k$, the actions of Q-learning are the actions defined in our LSM algorithm, and the reward function $r(\boldsymbol{x}_j, \boldsymbol{a}_j)$ is the expected number of stable queue users in (18). The content request distribution is predicted by the proposed LSM algorithm and the contents to be cached are determined by Theorem 1.
- The second baseline is the Q-learning algorithm in [34] without content caching, which we refer to as "Q-learning without cache". The settings of the Q-learning without content caching are similar to the Q-learning algorithm with content caching. However, in the Q-learning without content caching algorithm, all of the contents requested by the users are transmitted from the content server.

[1]The data is available at http://index.youku.com/.

TABLE II
SYSTEM PARAMETERS

| **Parameter** | **Value** | **Parameter** | **Value** | **Parameter** | **Value** |
|---|---|---|---|---|---|
| $P_K$ | 15 dBm | $L$ | 2 Mbits | $Z$ | 20 dB |
| $P_C$ | 20 dBm | $\eta^u_{\text{LoS}}, \eta^u_{\text{NLoS}}$ | 1.2, 23 | $\rho$ | 30 ms |
| $\sigma$ | -94 dBm | DIFS | 50 $\mu$s | $F_C$ | 100 Mbit |
| $F_l$ | 10 Mbit | $E[A]$ | 1500 bytes | $N_w$ | 2 |
| $F_u$ | 20 Mbit | $\delta$, $\delta^\alpha$ | 0.1, 0.05 | $\beta$ | 2 |
| $C$ | 3 | $W_1, W_2, W_3$ | 5,5,20 | ACK | 304 $\mu$s |
| CTS | 304 $\mu$s | $X, Y$ | 11.9, 0.13 | RTS | 352 $\mu$s |
| $N$ | 25 | $\eta^l_{\text{LoS}}, \eta^l_{\text{NLoS}}$ | 1, 20 | $\varsigma$ | 20 dB |
| $\varsigma_{\text{EE}}$ | 0.3 | $\varsigma_{\text{IE}}, \varsigma_{\text{II}}, \varsigma_{\text{EI}}$ | 0.2,0.1,0.4 | SIFS | 16 $\mu$s |
| $R_w$ | 4 Mbps | $W_1^\alpha, W_2^\alpha, W_3^\alpha$ | 5,5,30 | $\mathbb{P}_{\text{IN}}$ | 0.3 |
| $W$ | 2 | $S$ | 13.5 mV | $N_x$ | 4 |

We also compare our LSM based prediction algorithm with the echo state network (ESN) learning algorithm in [19]. All statistical results are averaged over 5000 independent runs.

In Fig. 4, we show how the average number of stable queue users changes as the number of UAVs varies. In this figure, the heuristic search for resource allocation algorithm represents a comparison algorithm that uses heuristic search to find optimal resource allocation and the proposed LSM algorithm for prediction and user association. From Fig. 4, we can see that the number of stable queue users increases as the number of UAVs increases. This is due to the fact that increasing the number of UAVs provides more connection options for the users, and, thus, improves the number of stable queue users. Fig. 4 also shows that the proposed LSM algorithm can yield up to 33.3% and 50.3% gains in terms of the number of stable queue users compared to Q-learning algorithm with cache and Q-learning without cache, respectively, for a network with 5 UAVs. These gains stem from the fact that the proposed LSM algorithm can use the historical user information to find an user association scheme and predict the users' content request distribution to improve content caching. From Fig. 4, we can also see that the proposed LSM algorithm can almost reach the same gain as the heuristic search algorithm which proves that the proposed search-based algorithm can find the optimal resource allocation for each UAV based on Theorem 1.

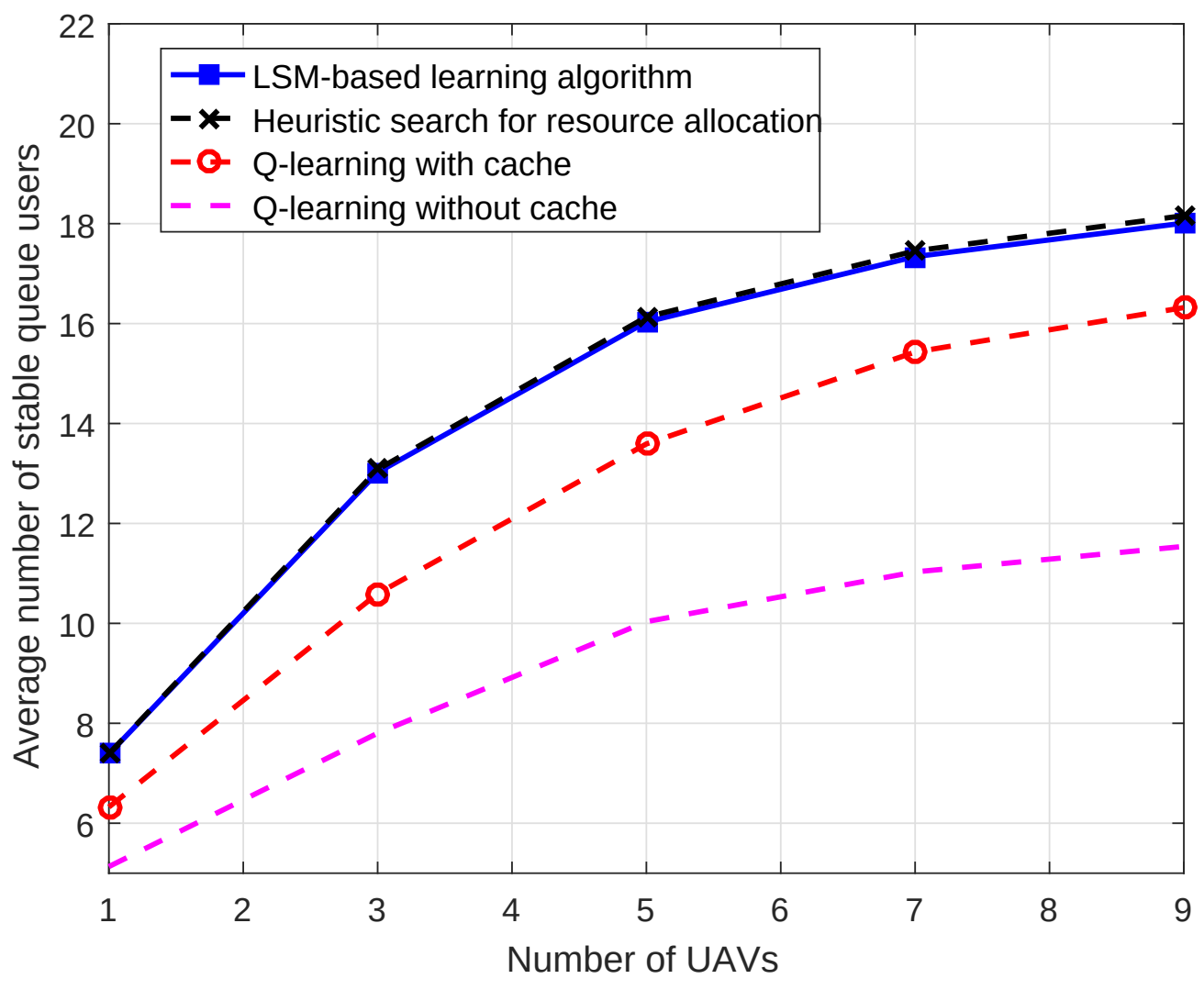

Fig. 4. Average number of stable queue users as the number of UAVs varies.

Fig. 5 shows the cumulative distribution function (CDF) for the data rate resulting from all of the considered schemes. In Fig. 5, we can see that the data rate of almost 100% of users resulting from all of the considered algorithms will be below 2 Mbps. This is due to the fact that, the data rate requirement of each user is 2 Mbps and, hence, whenever a given user's rate requirement is satisfied, the UAVs will allocate the licensed or unlicensed spectrums to other users. Fig. 5 also shows that the proposed approach improves the CDF of up to 27.56% and 48% gains at a rate of 1.8 Mbps compared to Q-learning with cache and Q-learning without cache, respectively. These gains stem from the fact that the proposed algorithm can store and use historical user information related to user association to find a better solution compared to Q-learning with cache and Q-learning without cache.

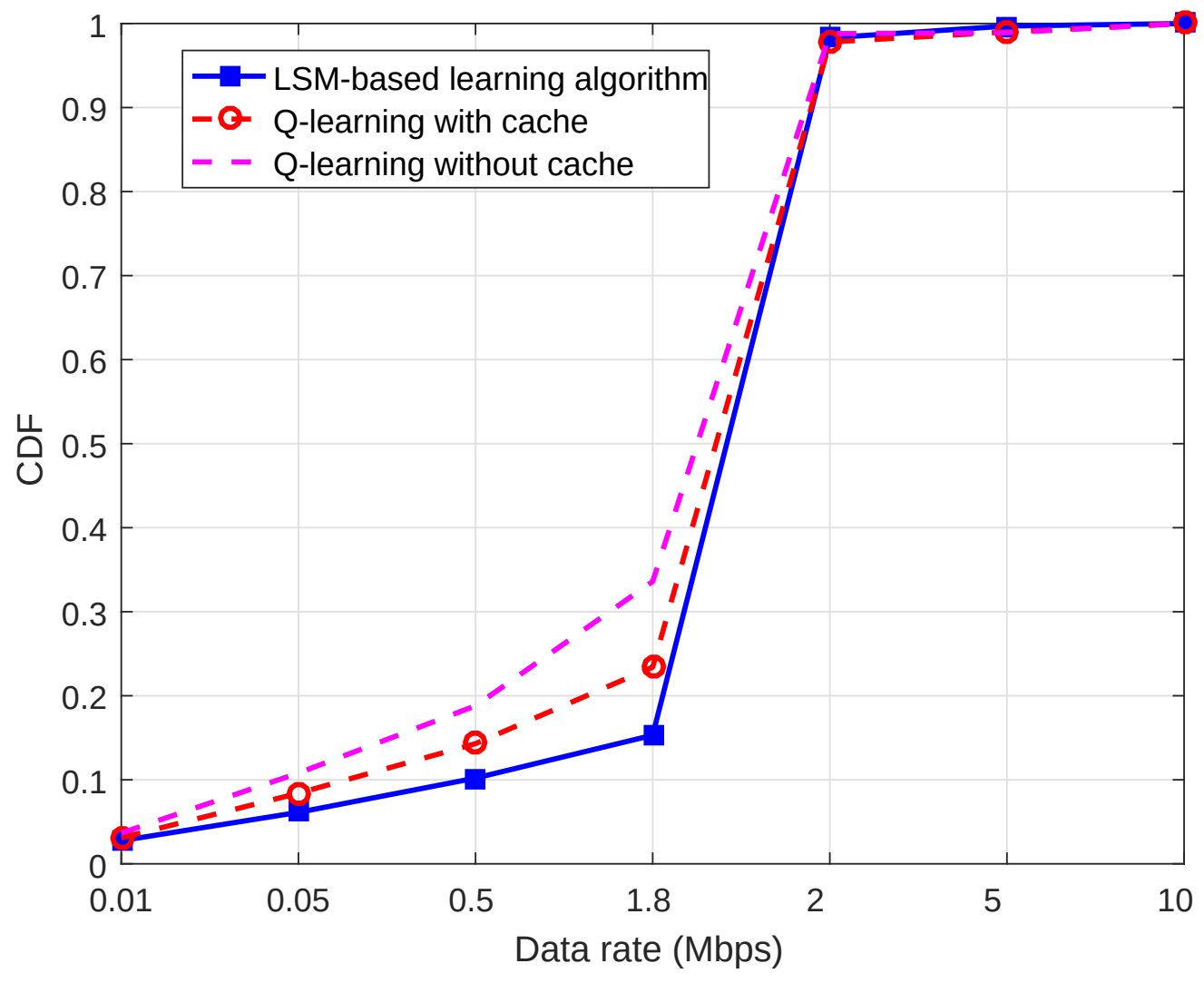

Fig. 5. CDFs of the data rate resulting from the different algorithms.

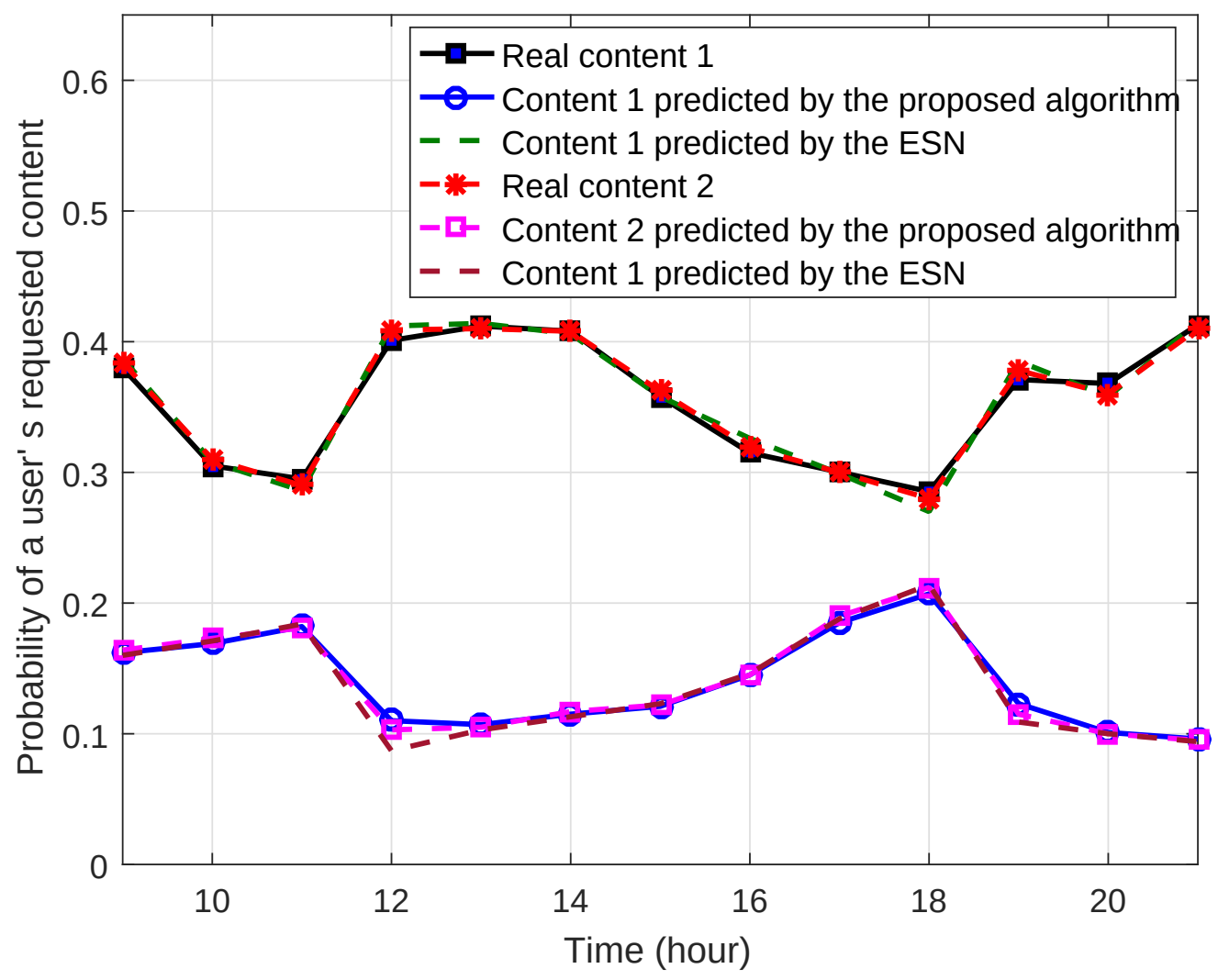

Fig. 6. Content request probability predictions.

In Fig. 6, we show the variations of two content request probabilities of an arbitrarily selected user during one day. From Fig. 6, we can see that the accuracy of the predictions of the proposed LSM algorithm is within less than 10% from the real content request probability. Since this gap does not affect the ranking of each content request probability, the cloud can find the optimal contents to cache using the proposed algorithm. From Fig. 6, we can also see that the proposed LSM algorithm can reduce 13.4% mean squared error compared to the ESN algorithm. This is due to the fact that the neurons in the LSM model are more dynamic than the neurons in the ESN. In consequence, the LSM model can store more information to predict the users content request distribution compared to the ESN model. Fig. 6 also shows that the sum of the probabilities with which this user requests contents 1 and 2 exceeds 0.5 during each hour. This is because the user always requests a small number of contents.

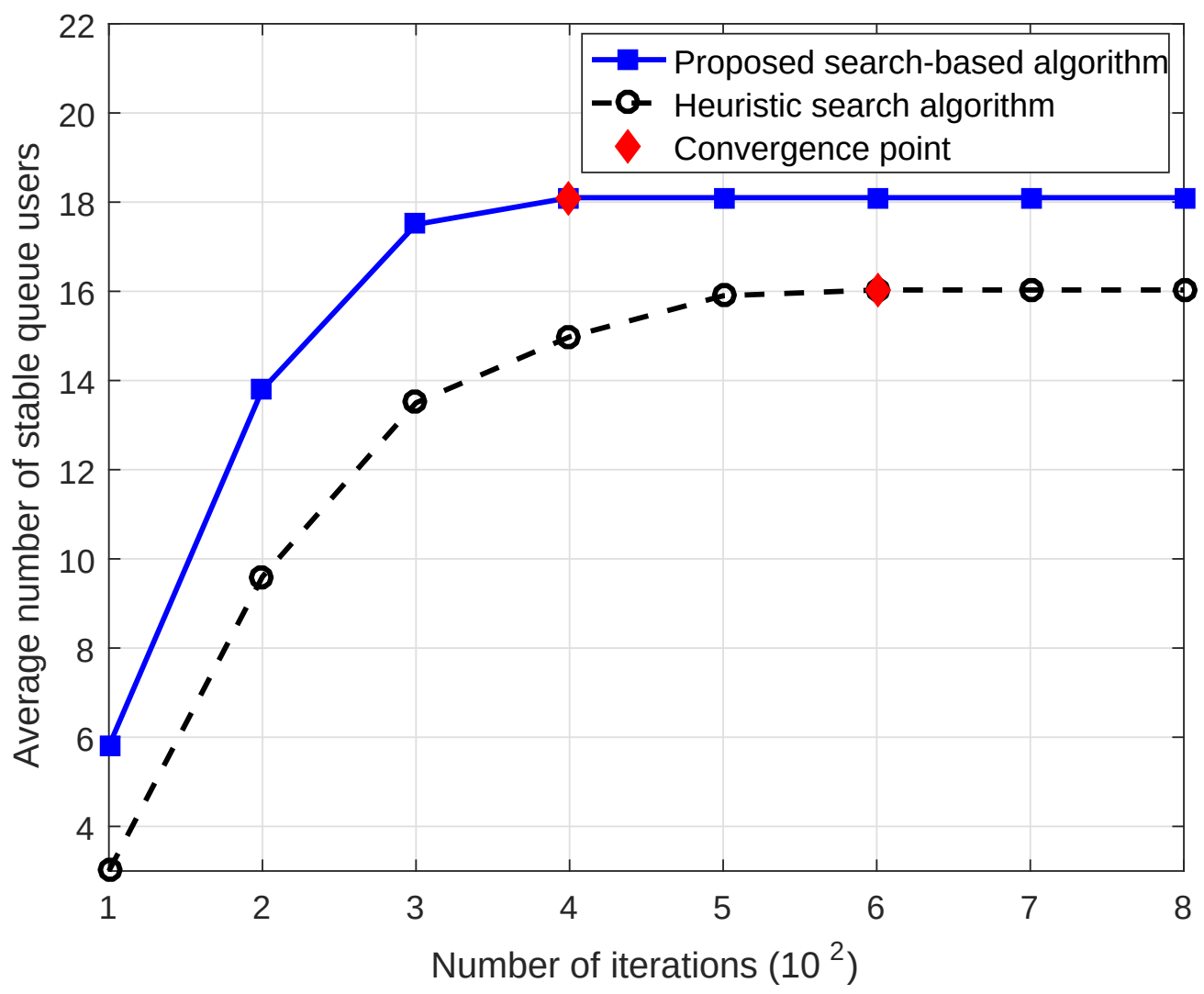

Fig. 7. Convergence of the learning algorithms.

Fig. 7 shows the number of iterations needed till convergence for both the proposed approach and Q-learning with cache. In this figure, we can see that, as time elapses, the number of stable queue users increases until convergence to their final values. Fig. 7 also shows that the proposed approach needs 400 iterations to reach convergence and exhibits a considerable reduction of 33.3% less iterations compared to Q-learning with cache. This is due to the fact that the LSM algorithm stores the users' states.

In Fig. 8, we show the variations in the number of the users that are allocated to the licensed and unlicensed bands as the number of contents stored at UAV cache varies. From Fig. 8, we can see that the number of users over the licensed and unlicensed bands increases as the number of cached contents increases. This is due to the fact that content caching offloads the traffic over the cloud-UAV links thus reducing the bandwidth needed for having stable queues at the users. Fig. 8 also shows that the number of users on the licensed band is 50% more than the number of users over the unlicensed band due to the difference in the parameters of the path loss over the unlicensed band.

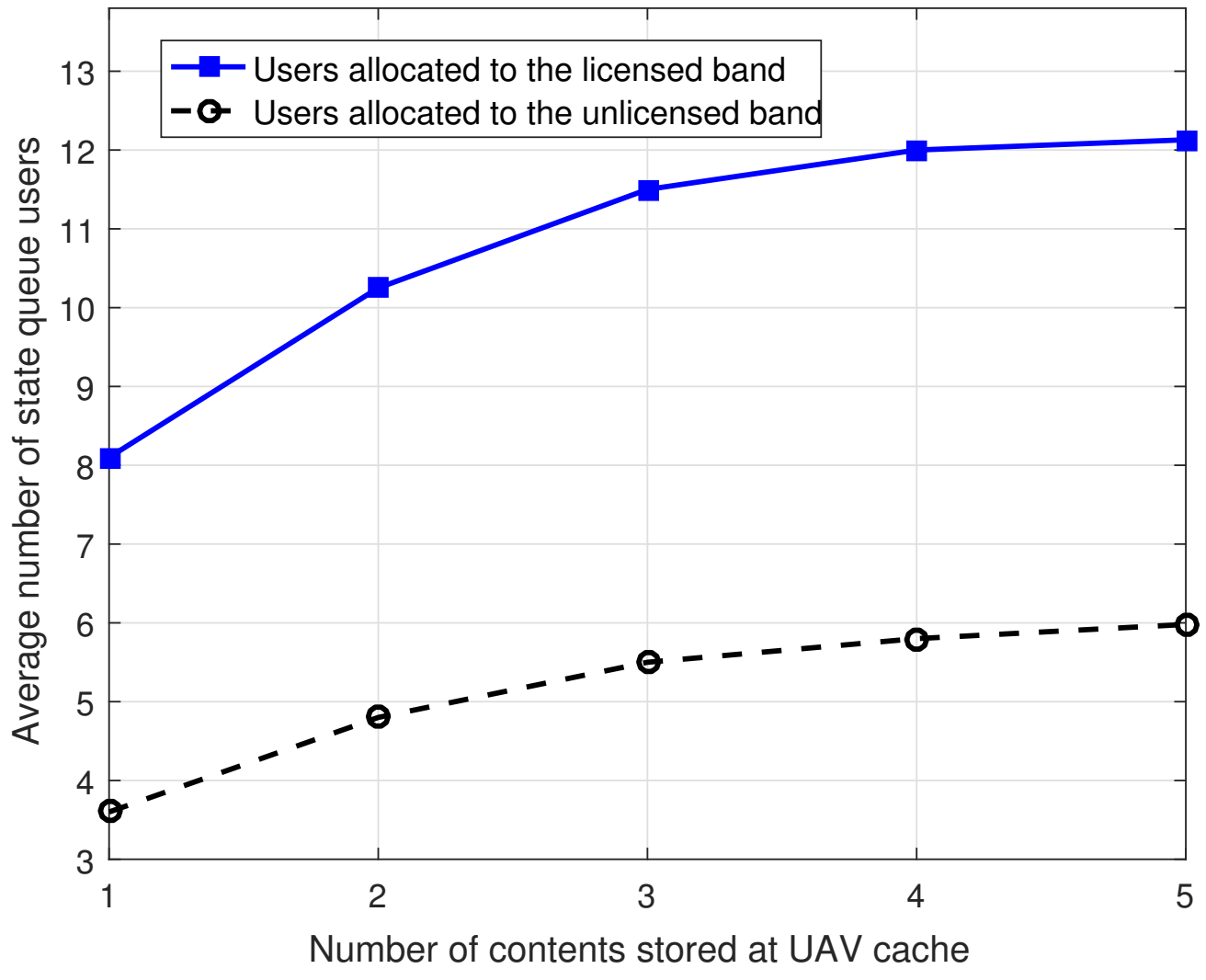

Fig. 8. Number of users associated with licensed and unlicensed bands as the number of stored contents at UAV cache varies.

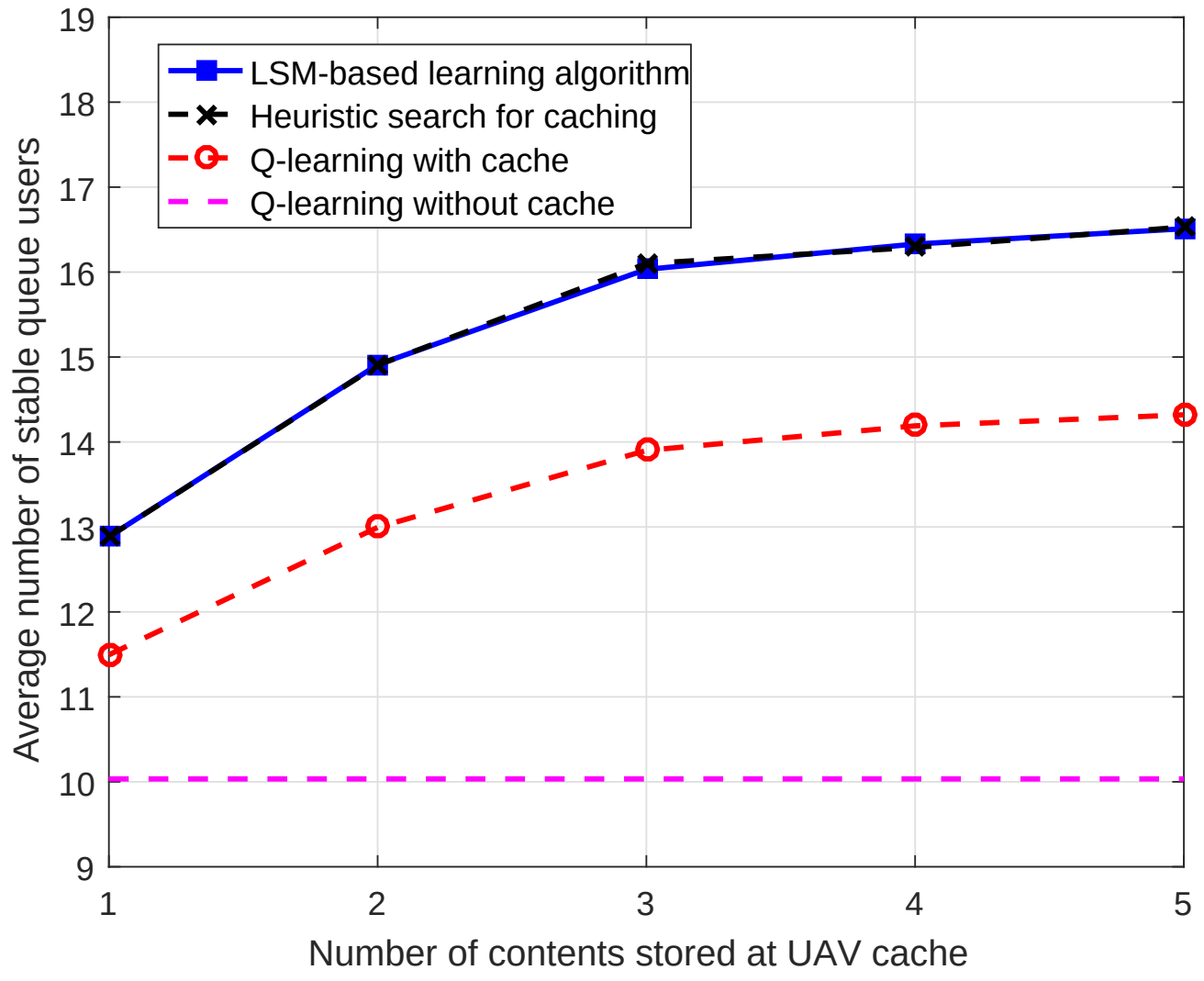

Fig. 9. Average number of stable queue users as the number of stored contents at UAV cache varies.

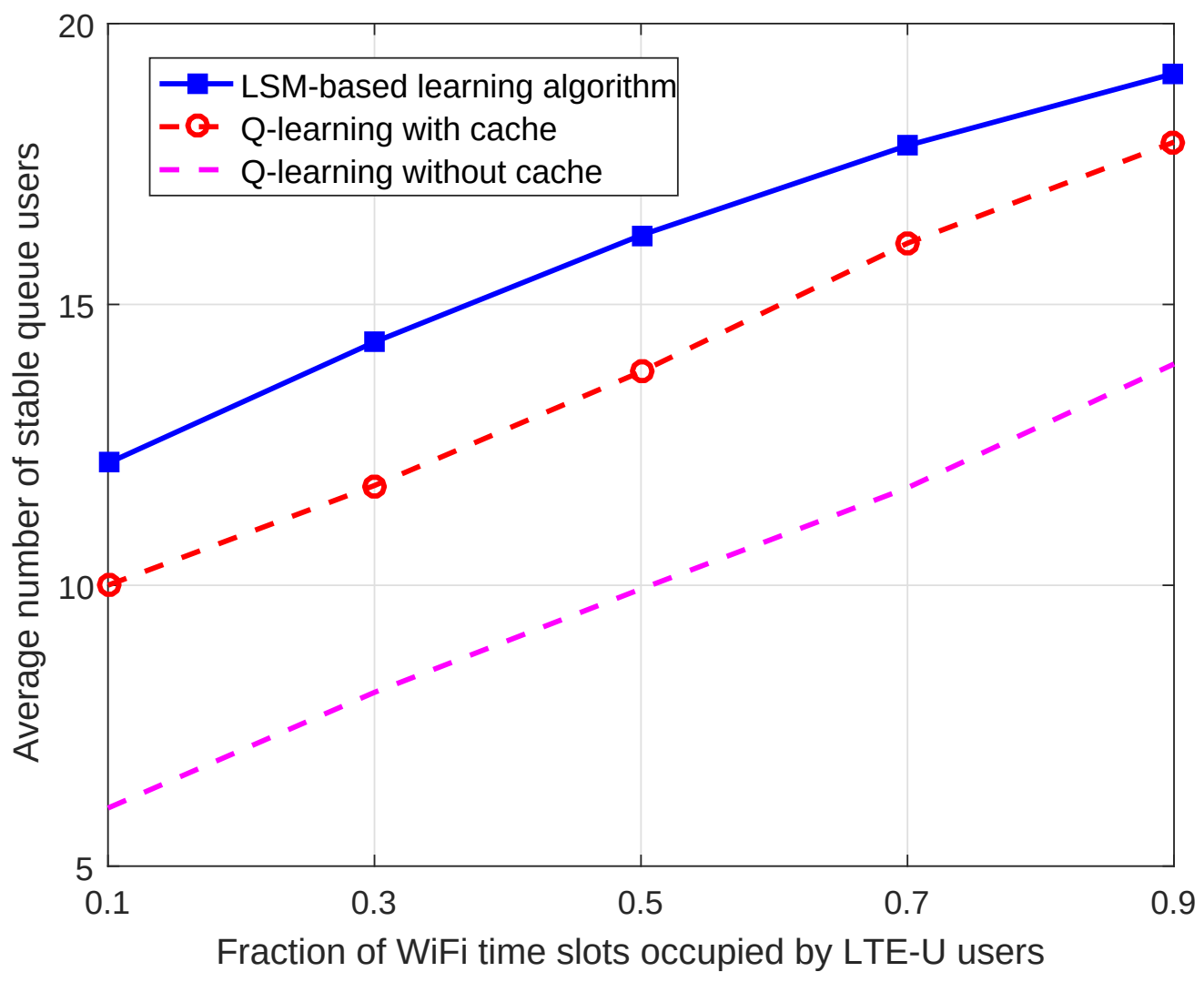

Fig. 10. Average number of stable queue users as the fraction of WiFi time slots occupied by LTE-U users varies.

Fig. 9 shows how the average number of stable queue users changes as the number of contents stored at UAV cache varies. In this figure, heuristic search for caching algorithm is a comparison algorithm that uses heuristic search to find the optimal caching contents and the proposed algorithm for prediction and user association. From Fig. 9, we can see that the average number of state queue users increases as the number of contents stored at UAV cache increases. This is because as the number of cached contents increases, the traffic load over the fronthaul can be reduced. From Fig. 9, we can also see that the proposed algorithm can achieve up to 15.5% and 64% gains in terms of the number of stable queue users compared to Q learning with cache and Q-learning without cache as the number of the contents stored at each UAV is 5. This is due to the fact that the LSM can store the users' historical information related to the user association and use this information to find the optimal user association. Fig. 9 also shows that the proposed LSM algorithm can reach almost the same performance in terms of the number of stable queue users as the heuristic search algorithm. This proves that the proposed Theorem 1 can find the optimal contents stored at the UAV cache.

In Fig. 10, we show how the average number of stable queue users changes as the fraction of WiFi time slots occupied by users varies. From Fig. 10, we can see that as the fraction of WiFi time slots occupied by users increases, the average number of the stable queue users increases. This stems from the fact that as the fraction of WiFi time slots occupied by users increases, each UAV can allocate more WiFi time slots to its associated users and, hence, each UAV can satisfy more users' data rate requirements. Fig. 10 also shows that the proposed algorithm can yield up to 12.5% and 35.7% gains in terms of the average number of stable queue users for a UAV can use 0.9 of WiFi time slots. The 12.5% gain is due to the fact that the proposed algorithm can use historical users information to find the optimal user association for each UAV. The 35.7% gain stems from the fact that the proposed algorithm can use caching technology to reduce the traffic load over the fronthaul and use historical information related to the user association to find the optimal user association.

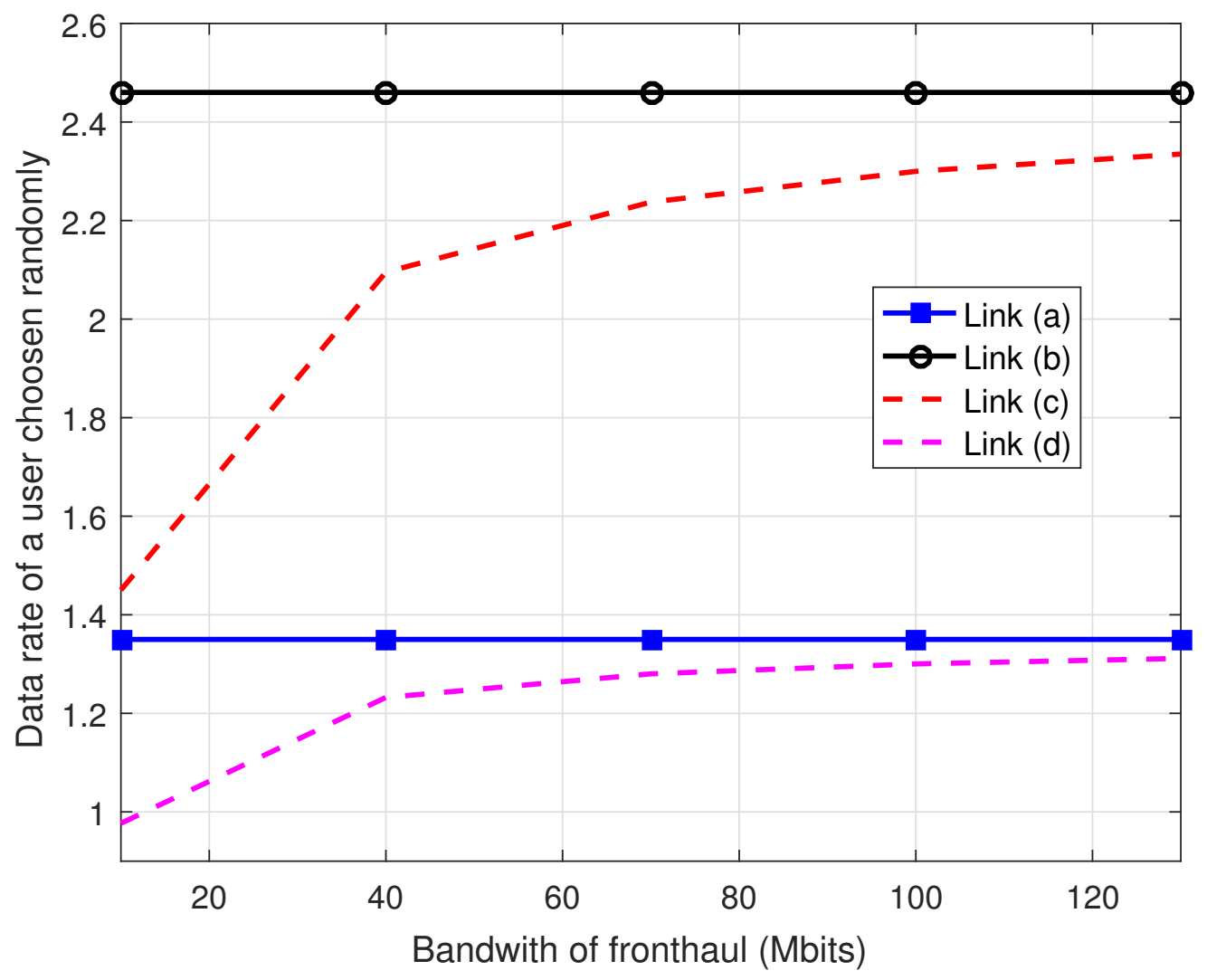

Fig. 11. Data rates of transmission links as the bandwidth of fronthaul varies.

Fig. 11 shows how the data rates of transmission links change as the bandwidth of the fronthaul varies. In this figure, the user is randomly chosen from the simulation. From Fig. 11, we can see that, as the bandwidth of the fronthaul (cloud-to-UAV links) increases, the data rates of links (a) and (b) remain unchanged while the data rates of links (c) and (d) increase. This is due to the fact that the contents transmitted over links (c) or (d) need to use the fronthaul. However, links (a) and (b) are used to directly transmit contents from the cache of any given UAV to a user without using the fronthaul. As the bandwidth of the fronthaul increases, the data rate of the fronthaul increases and, hence, the data rates of links (c) and (d) increase. Fig. 11 also shows that the unlicensed band data rate of links (b) and (c) are higher than the licensed band data rate of links (a) and (d). This stems from the fact that the user is interfered by the fronthaul

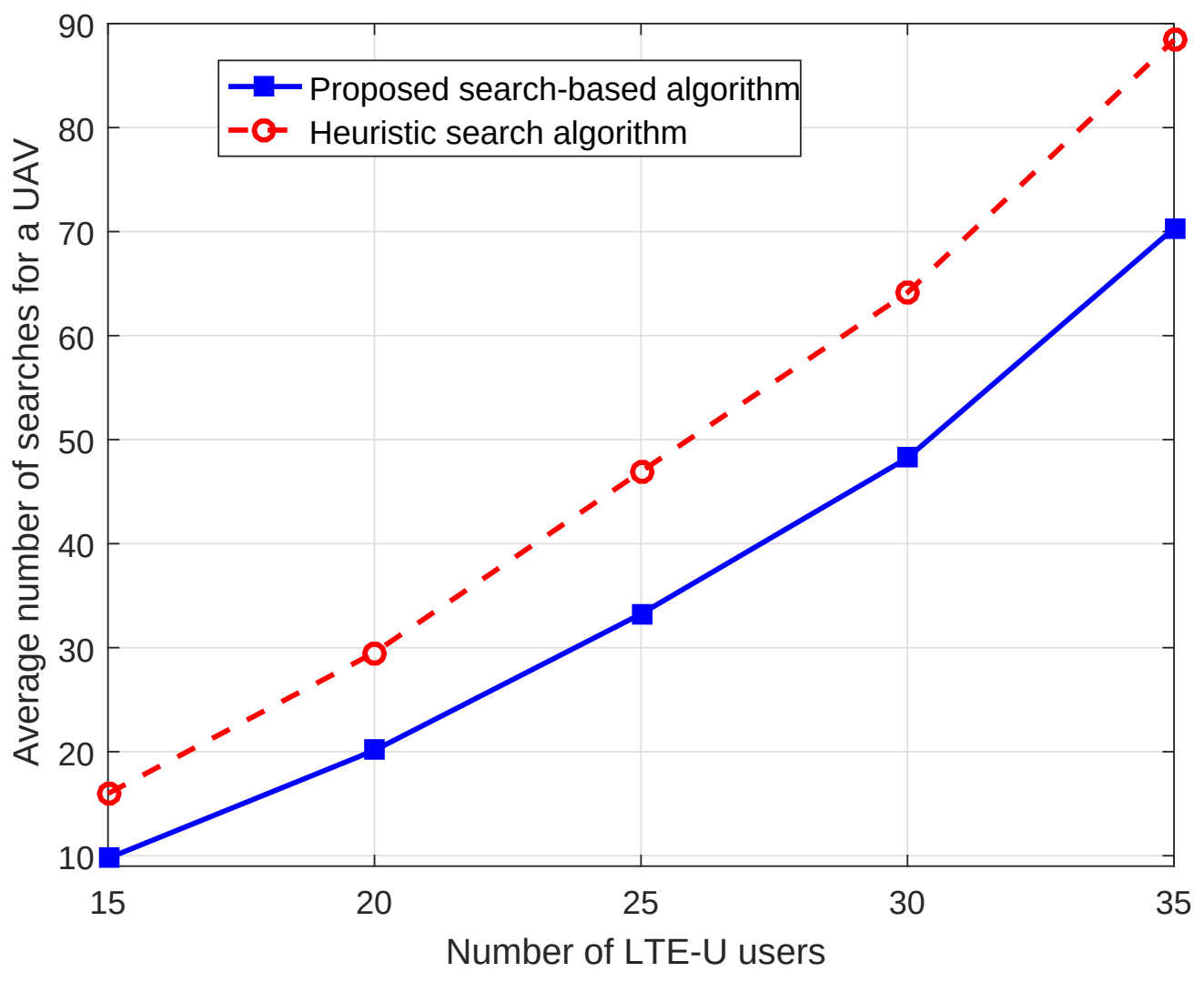

Fig. 12. Number of searches as the number of users changes.

transmission heavily.

In Fig. 12, we show how the average number of searches needed to find the optimal resource allocation varies as the number of users increases. From Fig. 12, we can see that, as the number of users increases, the average number of searches needed to find the optimal resource allocation for both algorithms increases. This is because as the number of the users increases, the number of the users associated with each UAV increases and, hence, each UAV needs to search more times to find the optimal resource allocation. Fig. 12 also shows that the proposed search-based algorithm can achieve up to 28.5% gain in terms of the number of searches needed to find the optimal resource allocation. This gain stems from the fact that the proposed search based algorithm can adjust the searching method based on the users' data rate requirements.

## V. Conclusion

In this paper, we have developed a novel framework that uses flying cache-enabled UAVs to provide service for users in an LTE-U system. We have formulated an optimization problem that seeks to maximize the number of stable queue users. To solve this problem, we have developed a novel algorithm based on the machine learning tools of liquid state network. The proposed prediction algorithm enables the cloud to predict each user's content request distribution and, thus, determine the UAV's cached contents. Using the proposed LSM resource allocation algorithm, each UAV can decide on its spectrum allocation scheme autonomously with limited information on the network state. Simulation results have shown that the proposed approach yields significant performance gains. Moreover, the results have also shown that the use of LSM can significantly improve convergence time when compared to Q-learning.